\documentclass[sigconf]{acmart}

\usepackage{xcolor}         
\usepackage[utf8]{inputenc} 
\usepackage[T1]{fontenc}    
\usepackage{hyperref}       
\usepackage{url}            
\usepackage{booktabs}       
\usepackage{amsfonts}       
\usepackage{nicefrac}       
\usepackage{microtype}      
\usepackage{amsmath}
\usepackage{tabularx}
\usepackage{graphicx}
\usepackage{multirow}
\usepackage{enumitem}
\usepackage{adjustbox}
\usepackage{algorithm}
\usepackage{algorithmic}
\usepackage[most]{tcolorbox}
\usepackage{caption}
\usepackage{xspace}
\usepackage{amsthm}
\newtheorem{proposition}{Proposition}

\AtBeginDocument{%
  }

\copyrightyear{2026}
\acmYear{2026}
\setcopyright{cc}
\setcctype{by}
\acmConference[KDD '26]{Proceedings of the 32nd ACM SIGKDD Conference on Knowledge Discovery and Data Mining V.2}{August 09--13, 2026}{Jeju Island, Republic of Korea}
\acmBooktitle{Proceedings of the 32nd ACM SIGKDD Conference on Knowledge Discovery and Data Mining V.2 (KDD '26), August 09--13, 2026, Jeju Island, Republic of Korea}
\acmDOI{10.1145/3770855.3817909}
\acmISBN{979-8-4007-2259-2/2026/08}

\acmSubmissionID{v2rtp1807}

\begin{document}

\title{Breaking Information Cocoons: A Hyperbolic Framework for Balancing Exploration and Exploitation in Recommender Systems}

\author{Qiyao Ma}
\affiliation{%
  \institution{University of California, Davis}
  \city{Davis}
  \country{United States}
}
\email{qiyma@ucdavis.edu}

\author{Menglin Yang}
\authornote{Corresponding author.}
\affiliation{%
  \institution{The Hong Kong University of Science and Technology (Guangzhou)}
  \city{Guangzhou}
  \country{China}
}
\email{menglinyang@hkust-gz.edu.cn}

\author{Mingxuan Ju}
\authornote{Authors affiliated with Snap Inc. served in advisory roles only for this work.}
\affiliation{%
  \institution{Snap Inc.}
  \city{Bellevue}
  \country{United States}
}
\email{mju@snap.com}

\author{Tong Zhao}
\authornotemark[2]
\affiliation{%
  \institution{Snap Inc.}
  \city{Bellevue}
  \country{United States}
}
\email{tong@snap.com}

\author{Neil Shah}
\authornotemark[2]
\affiliation{%
  \institution{Snap Inc.}
  \city{Bellevue}
  \country{United States}
}
\email{nshah@snap.com}

\author{Rex Ying}
\affiliation{%
  \institution{Yale University}
  \city{New Haven}
  \country{United States}
}
\email{rex.ying@yale.edu}

\def\model{HERec\xspace}
\begin{abstract}
    Modern recommender systems often create information cocoons, restricting users' exposure to diverse content. The central challenge is to balance content exploration and exploitation while allowing users to adjust their recommendation preferences. Ideally, this balance can be captured with a hierarchical representation, where depth search facilitates exploitation and breadth search enables exploration. However, existing approaches face two fundamental limitations: Euclidean methods struggle to capture hierarchical structures, while hyperbolic methods, despite their superior hierarchical modeling, lack semantic understanding of user and item profiles and fail to provide a principled mechanism for balancing exploration and exploitation. To address these challenges, we propose HERec, a hyperbolic framework that effectively balances exploration and exploitation in recommender systems. Our framework introduces two key innovations: (1) a semantic-enhanced hierarchical mechanism that aligns rich textual descriptions with collaborative information directly in hyperbolic space. Theoretical gradient analysis demonstrates that this alignment effectively leverages the underlying hyperbolic manifold structure, resulting in more accurate modeling of users and items; (2) an automatic hierarchical clustering mechanism by optimizing Dasgupta’s cost, which discovers hierarchical structures without requiring predefined hyperparameters, enabling user-adjustable exploration-exploitation trade-offs. Extensive experiments demonstrate that HERec consistently outperforms both Euclidean and hyperbolic baselines, achieving up to 5.49\% improvement in utility metrics and 11.39\% increase in diversity metrics, effectively mitigating information cocoons.
\end{abstract}

\ccsdesc[500]{Information systems~Recommender systems}
\ccsdesc[500]{Information systems~Personalization}
\ccsdesc[500]{Computing methodologies~Learning paradigms}

\keywords{Recommender Systems, Geometric Topology, Hyperbolic Space, Graph Neural Networks, Large Language Models}
\maketitle

\section{INTRODUCTION} 

\begin{figure}[t]
    \centering
    \includegraphics[width=0.8\columnwidth]{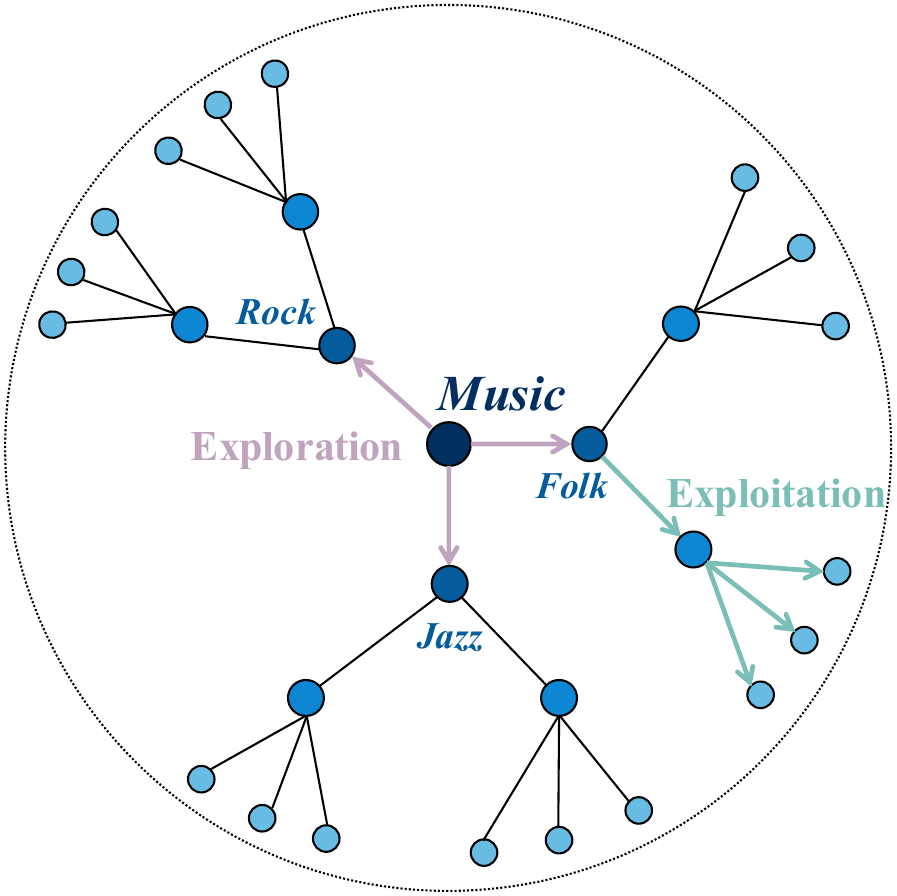}
    \caption{Exploration and exploitation in hyperbolic space.}
    \label{fig:Exp}
\end{figure}

As online platforms grow in size dramatically, users are overwhelmed by an influx of information, creating information cocoons that restrict their exposure to diverse and novel content~\cite{kunaver2017diversity, castells2021novelty}.

Traditional recommendation approaches have primarily relied on collaborative filtering and interaction data~\cite{he2020lightgcn, koren2009matrix}, where users with similar historical preferences are expected to share future interests~\cite{chen2021neural}. However, these methods face two critical challenges in improving user experience. First, existing models struggle to capture the underlying hierarchical structure inherent in user-item networks~\cite{zheng2017hierarchical, unger2020hierarchical}, a critical factor for enhancing performance and personalization. Second, these models fail to effectively balance the exploration-exploitation trade-off. For instance, a jazz enthusiast may want to explore rock music, but current models primarily focus on exploitation—recommending only familiar genres—without adequately supporting exploration.

Hyperbolic space has emerged as a promising approach for modeling hierarchical structures and power-law distributions~\cite{krioukov2010hyperbolic,yang2022hrcf,sun2021hgcf,yang2022hicf}, where its exponential volume growth naturally aligns with hierarchical, scale-free structures~\cite{ganea2018hyperbolic, chami2019hyperbolic, liu2019hyperbolic, zhang2021lorentzian, yang2023kappahgcn}. In hyperbolic space, distances increase exponentially from the origin, allowing popular items or exploratory users to be positioned closer to the origin, while niche items or users with focused interests spread toward the boundaries. This spatial configuration enables hyperbolic graph neural network (GNN) models to perceive hierarchical relationships effectively, enhancing recommendation performance~\cite{yang2022hicf, sun2021hgcf, yang2022hrcf}. An illustrative example is shown in Figure~\ref{fig:Exp}.

Despite these advantages, current hyperbolic recommender systems still face key limitations. First, they struggle with a lack of semantic understanding of user and item profiles. While large language models (LLMs) offer a potential solution, their text encoders operate in Euclidean space, making it challenging to integrate LLM-derived semantic features with hyperbolic collaborative signals. This mismatch leaves existing hyperbolic models vulnerable to noise in collaborative information, as they fail to capture semantic insights effectively. Second, hyperbolic recommender systems lack a principled approach to balancing exploration (introducing new content) and exploitation (focusing on existing preferences)~\cite{wilson2021balancing}. This limitation reduces their ability to adapt to the exploration-exploitation trade-off, ultimately diminishing both the utility and diversity of recommendations.

\noindent \textbf{Proposed Method.} To address these challenges, we propose \model, a hyperbolic framework designed to balance exploration and exploitation in recommender systems. Our approach introduces two key innovations: (1) a hierarchical-aware semantic-collaborative alignment mechanism, which jointly aligns textual descriptions with user-item collaborative information in hyperbolic space, \textbf{supported by theoretical gradient analysis demonstrating its adaptiveness}; (2) a novel hierarchical representation structure that enables user-adjustable exploration-exploitation trade-offs. Unlike conventional approaches that require manually defined hierarchies, we propose a hyperparameter-free clustering mechanism theoretically optimized by Dasgupta’s cost~\cite{dasgupta2016cost}. This approach \textbf{automatically discovers hierarchical structures without requiring predefined hyperparameters}, allowing for a more adaptive and principled organization of user preferences.

To validate the effectiveness of \model, we conducted extensive experiments across multiple dimensions. Our model consistently outperformed baselines in both utility and diversity, marking a significant achievement as a single model to excel in both aspects simultaneously. For a fair comparison, we also enhanced Euclidean baselines with feature information, yet our model still demonstrated superior performance. Additionally, we conducted a detailed analysis on head and tail items, showing that \model effectively boosts recommendations for tail items, thus enhancing diversity. An ablation study, along with a hierarchical representation structure analysis module, confirms \model's capability to leverage both semantic meaning and collaborative information, providing deeper insights into the advantages of our approach.

\noindent \textbf{Contributions.} Our primary contributions focus on three key aspects: capturing collaborative signals, integrating semantic information, and modeling the underlying hierarchical structure. The proposed hyperbolic alignment framework unifies collaborative and semantic information in hyperbolic space, while a hierarchical representation structure organizes user preferences, allowing for dynamic adjustment between exploration and exploitation. Moreover, through extensive experiments evaluating both utility and diversity metrics, we demonstrate that \model achieves state-of-the-art performance across utility and diversity metrics.
We have made our code available at: \textcolor{cyan}{~\url{https://github.com/Martin-qyma/HERec}}.
\section{Related Work}
\label{sec:relate}
In this section, we review graph-based collaborative filtering, large language models (LLMs) for recommendation, hyperbolic representation learning methods and existing approaches that balance the trade-off between utility and diversity.

\subsection{Graph Collaborative Filtering}
Graph Collaborative Filtering has emerged as a dominant approach in recommender systems, with graph neural networks (GNNs) playing a central role. Unlike traditional collaborative filtering techniques, such as matrix factorization~\cite{koren2009matrix}, GNNs capitalize on the relational structure in user-item interactions, enabling a more nuanced understanding of complex dependencies. Prominent graph-based methods, such as NGCF~\cite{wang2019neural}, SGL~\cite{wu2021self}, and LightGCN~\cite{he2020lightgcn}, have been widely adopted and have consistently achieved state-of-the-art performance. Nevertheless, these models are constrained to Euclidean space, which may fall short in capturing the underlying power-law distribution in user-item networks.

\subsection{LLMs for Recommendation}
Large language models (LLMs) are increasingly used in recommender systems for their strong capability in understanding textual information. Some approaches leverage LLMs directly for downstream recommendation tasks~\cite{bao2023tallrec, li2023text, lyu2023llm, zheng2024adapting, zhang2023collm}, while others utilize LLMs as an auxiliary tool for collaborative filtering~\cite{ren2024representation, xi2024towards, wei2024llmrec}. However, these approaches require substantial inference time and are computationally inefficient. In our \model, we use language models exclusively to improve data quality, eliminating the need for repeated inference during each recommendation by utilizing hyperbolic embeddings for the final output.



\subsection{Hyperbolic Learning and Recommendation}
Recent research has explored hyperbolic space as a robust alternative to Euclidean geometry for modeling the hierarchical and non-Euclidean structures inherent in user-item interactions~\cite{chami2019hyperbolic, liu2019hyperbolic, yang2022hrcf, zhang2021lorentzian}. By leveraging the property that hyperbolic volume grows exponentially with the radius, these models effectively capture the power-law distributions and latent hierarchies typical of scale-free graphs~\cite{chen2022modeling, ma2021knowledge, ying2018graph}. Key advancements include the development of fully hyperbolic graph convolutions to minimize structural distortion~\cite{wang2021fully} and manifold-aware approaches for knowledge-aware and social recommendation~\cite{sun2021hgcf, yang2022hicf, zhang2022hyperbolic}. Collectively, these works demonstrate that hyperbolic manifolds provide superior representation capacity for graph-structured data compared to traditional Euclidean methods.

However, none of these hyperbolic models fully leverage the rich semantic information embedded in language-based datasets. While some approaches incorporate knowledge graphs as supplementary information~\cite{tai2021knowledge, du2022hakg}, these models primarily focus on refining hierarchical structures rather than achieving a deeper semantic understanding of each item or user.

\subsection{Utility-Diversity Trade-off}
Various strategies have been proposed to balance utility and diversity. Some leverage reinforcement learning to foster exploration~\cite{li2023break, liu2021diversity}, while others mitigate data bias to optimize trade-offs~\cite{Luo2023cross} or integrate fairness objectives directly into training~\cite{Masrour_Wilson_Yan_Tan_Esfahanian_2020, wu2022multi}. Notably, recent work suggests that the accuracy-diversity tension may not be inherent, but rather a byproduct of standard metrics and user information cocoons~\cite{peng2024reconciling}.

Despite these advances, existing methods still lack a systematic framework for explicitly modeling the utility-diversity trade-off and its underlying structure. While these approaches recognize the heterogeneity among users and items, such as the ``long tail'' phenomenon, they do not provide principled mechanisms to capture or leverage this structure. Consequently, prior work has largely focused on increasing diversity, rather than systematically breaking information cocoons through a well-founded trade-off strategy.
\section{PRELIMINARIES}
\label{sec:preliminary}
Embeddings learned from graph collaborative filtering are by default in the Euclidean space, which is unable to capture hierarchical structures. To address the problem, we introduce how to derive these embeddings in a hyperbolic space.

\noindent \textbf{Riemannian Manifold.} A Riemannian manifold $(\mathcal{M}, g)$ is a smooth manifold with an inner product $g_x$ at each point $x \in \mathcal{M}$, defining geometric properties such as angles and curve lengths. Its curvature determines the geometry: positive (elliptic), zero (Euclidean), or negative (hyperbolic). We focus on hyperbolic geometry, modeled using the Lorentz (hyperboloid) representation.

\noindent \textbf{Lorentz Model.} An $n$-dimensional Lorentz manifold with negative curvature $-1/\kappa$ $(\kappa>0)$ is described as the Riemannian manifold $(\mathbb{H}^n_\kappa, g_\mathcal{L})$. Here, $\mathbb{H}^n_\kappa = \{x \in \mathbb{R}^{n+1} : \langle x, x \rangle_\mathcal{L} = -\kappa, x_0 > 0\}$, $g_\mathcal{L} = \eta$ with $\eta = \mathbf{I}_n$ except $\eta_{0,0} = -1$, and $\langle \cdot, \cdot \rangle_\mathcal{L}$ represents the Lorentzian inner product, which is defined as:
\begin{equation}
    \langle x, y \rangle_\mathcal{L} := -x_0 y_0 + \sum_{i=1}^n x_i y_i
\end{equation}
A tangent space is an $n$-dimensional vector space that locally approximates hyperbolic space. For a point $x \in \mathbb{H}_\kappa^n$, the tangent space $\mathcal{T}_x\mathbb{H}_\kappa^n$ is defined as:
\begin{equation}
    \mathcal{T}_x\mathbb{H}_\kappa^n := \{v \in \mathbb{R}^{n+1} : \langle v, x \rangle_\mathcal{L} = 0\}
\end{equation}
This space is orthogonal to $x$ under the Lorentzian inner product, providing accurate local approximation. Transitions between hyperbolic space and tangent space are facilitated by exponential and logarithmic maps, essential for gradient-based optimization. The exponential map $\exp_x: \mathcal{T}_x\mathbb{H}_\kappa^n \to \mathbb{H}_\kappa^n$ projects a tangent vector onto hyperbolic space along a geodesic:
\begin{equation}
    \exp_{x}^{\kappa}(v) = \cosh \left( \frac{\|v\|_{\mathcal{L}}}{\sqrt{\kappa}} \right) x + \sqrt{\kappa} \sinh \left( \frac{\|v\|_\mathcal{L}}{\sqrt{\kappa}} \right) \frac{v}{\|v\|_{\mathcal{L}}}
    \label{eq:exp}
\end{equation}
where $\|v\|_\mathcal{L} = \sqrt{\langle v, v \rangle_\mathcal{L}}$ is the Lorentzian norm. The logarithmic map $\log_x$ is the inverse of $\exp_x$ and retrieves the tangent vector for a point $y \in \mathbb{H}_\kappa^n$:
\begin{equation}
    \log_{x}^{\kappa}(y) = d_{\mathcal{L}}^\kappa(x, y) \frac{y + \frac{1}{\kappa} \langle x, y \rangle_\mathcal{L} x}{\|y + \frac{1}{\kappa} \langle x, y \rangle_\mathcal{L} x\|}
\end{equation}
where $d_\mathcal{H}^\kappa(x, y)$ is the distance between $x$ and $y$ in $\mathbb{H}^n_\kappa$:
\begin{equation}
    d_\mathcal{H}^\kappa(x, y) = \sqrt{\kappa} \, \text{arcosh} \left( -\langle x, y \rangle_\mathcal{L}/\kappa \right)
\end{equation}
We use $\mathbf{o} := \{\sqrt{\kappa}, 0, \ldots, 0\} \in \mathbb{H}^n_\kappa$ as the reference point. For simplicity, we set $\kappa = 1$, giving a curvature of $-1$, and omit $\kappa$ in subsequent discussions.

\section{METHODOLOGY}
\label{sec:model}
\begin{figure*}[t]
    \centering
    \includegraphics[width=\textwidth]{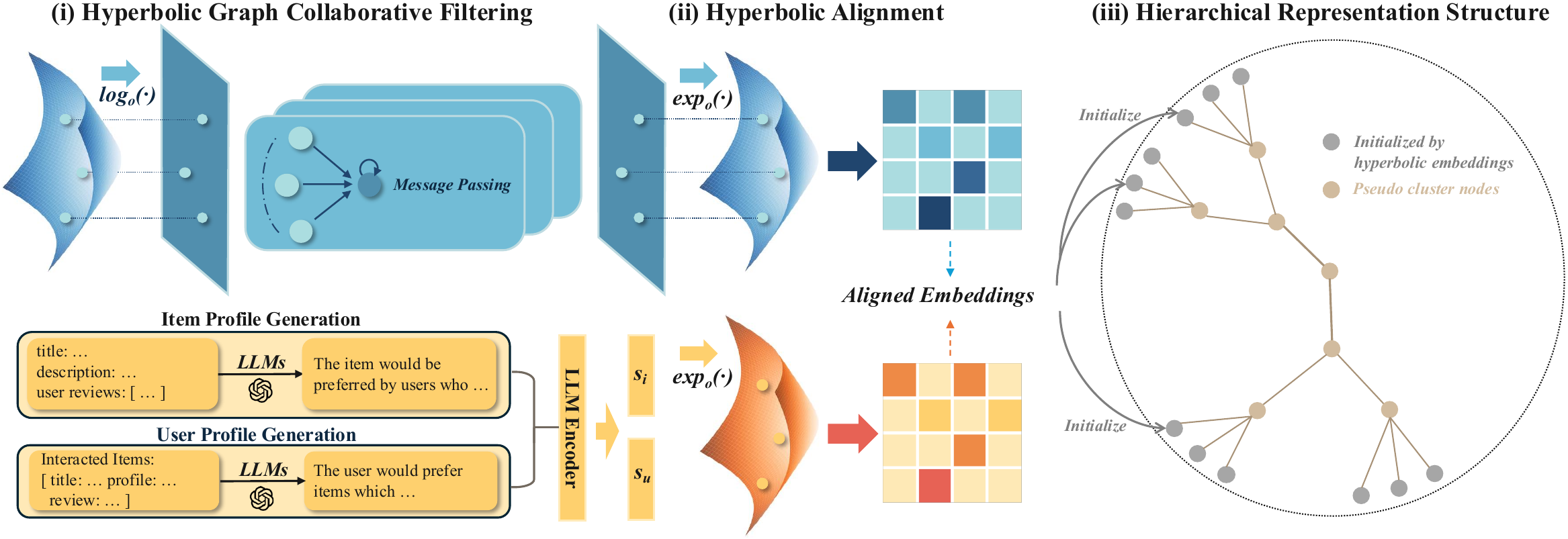}
    \vspace{-0.1in}
    \caption{The overall architecture of \model. (i) Hyperbolic Graph Collaborative Filtering: Encodes collaborative information using hyperbolic GNNs; (ii) Hyperbolic Alignment: Aligns semantic and collaborative information within hyperbolic space; (iii) Hierarchical Representation Structure: Builds hierarchy structure from hyperbolic embeddings.}
    \label{fig:framework}
    \vspace{-0.1in}
\end{figure*}

In this section, we present a comprehensive overview of our proposed model, \model. Our approach not only aims to deliver accurate recommendations but also addresses users' growing demand for more diverse suggestions. Furthermore, by constructing a hierarchical tree from the learned embeddings, we introduce an additional mechanism that enables users to control the degree of exploration in the recommended items. The overall framework is in Figure \ref{fig:framework}.

\subsection{Hyperbolic Graph Collaborative Filtering}
\label{subsec:Hyperbolic Graph Collaborative Filtering}
Similar to Euclidean space, hyperbolic graph collaborative filtering captures user-item dependencies via message passing mechanism. We apply the Lorentz representation for both users and items. 

\noindent \textit{Hyperbolic embedding initialization.} As discussed in section 3, we fix the origin $\mathbf{o} := \{\sqrt{\kappa}, 0, \ldots, 0\} \in \mathbb{H}^n_\kappa$ and use it as the reference point. Embeddings are initialized in Euclidean space using Gaussian sampling and projected onto the tangent space of $\mathbf{o}$ to derive their hyperbolic representations:
\begin{equation}
\begin{aligned}
    z_u^0 = (0, e_u), \quad\qquad   z_i^0 &= (0, e_i) \\
    h_u^0 = \exp_\mathbf{o}(z_u^0), \quad\quad h_i^0 &= \exp_\mathbf{o}(z_i^0)
\end{aligned}
    \label{equ: initialization}
\end{equation}
where $e_u, e_i$ are Gaussian-initialized Euclidean embeddings, and $z_u^0, z_i^0$ are their tangent space projections. $h_u^0, h_i^0$ denote the resulting hyperbolic embeddings.

\noindent \textit{Hyperbolic Message Passing.} User-item interactions are captured via hyperbolic neighborhood aggregation. For user $u$, item $i$, and their neighborhoods $\mathcal{N}_u, \mathcal{N}_i$, tangent space embeddings at layer $l$ are updated as:
\begin{equation}
    z_u^{l} = z_u^{l-1} + \sum_{i\in \mathcal{N}_u}\frac{1}{|\mathcal{N}_u|}z_i^{l-1},\quad
    z_i^{l} = z_i^{l-1} + \sum_{u\in \mathcal{N}_i}\frac{1}{|\mathcal{N}_i|}z_u^{l-1}
\end{equation}
where $|\mathcal{N}_u|$ and $|\mathcal{N}_i|$ are the neighborhood sizes. The final tangent embeddings are obtained by summing over all layers:
\vskip -0.1in
\begin{equation}
    z_u = \sum_{l} z_u^l, \quad\quad z_i = \sum_{l} z_i^l
\end{equation}
\vskip -0.1in
Since $z_u$ and $z_i$ are on the tangent space, we still need to project them into hyperbolic space to derive the hyperbolic embeddings:
\begin{equation}
    h_u=\exp_\mathbf{o}(z_u), \quad\quad h_i=\exp_\mathbf{o}(z_i)
\end{equation}

\noindent \textit{Hyperbolic Prediction Function.} Through hyperbolic messaging passing, we've captured high-order relational information into hyperbolic embeddings. Hyperbolic distance is now used to define a prediction function, which predicts the possibility that a user would interact with an item: 
\begin{equation}
    p(u,i) = \frac{1}{d_\mathcal{H}(h_u,h_i)}
\end{equation}

\noindent \textit{Hyperbolic Margin Ranking and Informative Mechanism.} To address the limitations of Euclidean-based approaches, we adopt the hyperbolic margin ranking loss~\cite{sun2021hgcf}, an extension of Bayesian Personalized Ranking (BPR)~\cite{rendle2012bpr}. This loss pulls user-item pairs with observed interactions closer while pushing negative pairs apart, preventing the collapse seen in Euclidean space by incorporating a margin:
\begin{equation}
    \ell_{m}(u,i,j) = \max(d_\mathcal{H}^2(h_u, h_i) - d_\mathcal{H}^2(h_u, h_j) + m, 0)
\end{equation}
where $d_\mathcal{H}$ represents hyperbolic distance, $m$ is the margin, $(u,i)$ is an observed interaction, and $(u,j)$ is a sampled negative pair.

To further refine this pull-push process, we integrate a geometric-aware mechanism~\cite{yang2022hicf} combining Hyperbolic-aware Margin Learning (HAML) and Hyperbolic Informative Negative Sampling (HINS). HAML dynamically adjusts the margin based on hyperbolic geometry, assigning a larger margin to pairs near the origin:
\begin{equation}
    \begin{aligned}
        m^\mathcal{H}_{ui} &= \mathrm{sigmoid}(\delta)\\
        \delta &= \frac{d_\mathcal{H}^2(e_u, \mathbf{o}) + d_\mathcal{H}^2(e_i, \mathbf{o}) - d_\mathcal{H}^2(e_u, e_i)}{e_{u,0} \cdot e_{i,0}}
    \end{aligned}
\end{equation}
where $e_{u,0}$ and $e_{i,0}$ are the zeroth coordinates of $e_u$ and $e_i$, representing their norms. HINS enhances negative sampling by prioritizing candidates near the positive item, leveraging hyperbolic geometry to better model head and tail items, ultimately improving recommendation performance.

\subsection{Semantic Representations Generation}
\label{subsec: Semantic Representations Generation}
We propose a paradigm to distill useful information from raw text descriptions of items and users into dense semantic embeddings for our alignment framework. Building on prior work~\cite{ma2024xrec, ren2024representation} that leverages large language models (LLMs) to extract meaningful user/item descriptions from noisy textual data, we generate rich user/item profiles as $\mathcal{P}_i = \textit{LLMs}(\mathcal{S}_i, \mathcal{T}_i)$, where $\mathcal{S}_i$ are system prompts, $\mathcal{T}_i$ is raw textual input, and $\mathcal{P}_i$ contains sentence-level descriptions.

To integrate these profiles with hyperbolic collaborative filtering embeddings, we encode the textual information into embeddings using pre-trained text encoders: $e_i = \textit{Encoder}(\mathcal{P}_i)$. While LLM embeddings are an alternative, they often incur higher computational overhead and are less optimized for alignment with hyperbolic representations, making pre-trained encoders a more practical choice.

\subsection{Hyperbolic Alignment}
We propose a hyperbolic alignment framework that unifies collaborative signals from hyperbolic embeddings with semantic cues from textual embeddings, producing a single representation that integrates both sources of knowledge. To achieve this, we develop a hyperbolic alignment loss, which reduces noise in representations and improves tail item recommendations by aligning information in the same hyperbolic space. A theoretical advantage of hyperbolic alignment over Euclidean alignment is discussed in the next subsection.

Before calculating the alignment loss, two steps are necessary: 1) the semantic embeddings must be adjusted to match the dimensionality of the collaborative embeddings, and 2) they should be transformed into a common representation space. To accomplish this, we first apply a multi-layer perceptron (MLP) to the semantic embeddings to ensure dimensional alignment. We then project the semantic embeddings into hyperbolic space so they can align with the hyperbolic collaborative embeddings. The hyperbolic alignment loss is calculated as: 
\begin{equation}
    \ell_{align}(i) = d_\mathcal{H}^2(h_i, s_i)
\end{equation}
where $d_{\mathcal{H}}$ represents the hyperbolic distance, $h_i$ is the hyperbolic embedding with collaborative information and $s_i$ denotes the projected semantic embedding in hyperbolic space.


\subsection{Gradient Analysis of Hyperbolic Alignment}
\label{sub:gradient analysis}
To demonstrate the unique advantage of hyperbolic alignment, we analyze the gradient behavior with respect to nodes of varying norms in Proposition~\ref{prop}, detailed proof with analysis can be found in Appendix \ref{sec:proof}. This provides insight into the hierarchical preservation properties of the hyperbolic space, which are absent in Euclidean space.

\begin{proposition}
Let $\mathbf{x}, \mathbf{y}$ denote embeddings in hyperbolic space $\mathcal{H}^n$ with corresponding unit direction vectors in space-like dimension $\hat{\mathbf{x}}, \hat{\mathbf{y}}$, and let $\theta$ be the angle between them. The gradient magnitude of semantic alignment satisfies:
\begin{equation}
    \|\nabla_{\mathbf{x}} d_{\mathcal{H}}(\mathbf{x}, \mathbf{y})\| \approx \frac{\|\hat{\mathbf{x}} - \hat{\mathbf{y}}\|}{\|\mathbf{x}\|(1 - \cos \theta)}.
\end{equation}
This facilitates adaptive gradient updates that intrinsically preserve hierarchical structures. In contrast, Euclidean space exhibits constant gradient magnitude $\|\nabla_{\mathbf{x}} d_{\mathcal{E}}(\mathbf{x}, \mathbf{y})\| = 1$, where $d_{\mathcal{H}}$ and $d_{\mathcal{E}}$ denote hyperbolic and Euclidean distances respectively.
\label{prop}
\end{proposition}

This proposition highlights that hyperbolic space offers adaptive gradient magnitudes based on node norms. Nodes with large norms (fine-grained preferences, small positional changes cause significant distance changes) receive smaller gradient updates, ensuring precise distance adjustments and preserving local structures. Meanwhile, nodes with smaller norms (abstract preferences, positional changes have less impact on distances) adjust more, refining global relationships and effectively capturing the hierarchical structure. This inherent ability of hyperbolic space to modulate learning intensity based on an embedding's position within the hierarchy offers a distinct advantage over the fixed-update approach of Euclidean geometry, ultimately facilitating more faithful and structured representations.

\subsection{Hierarchical Representation Structure}
\label{subsec: Hierarchy Structure}

In this subsection, we design a mechanism that allows users to balance the exploration-exploitation trade-off by constructing a hierarchical representation structure based on the learned hyperbolic embeddings.

\subsubsection{Hierarchy Tree}
The primary motivation for constructing a hierarchy tree is to uncover hidden structures in the data that are not explicitly reflected in the dataset. For instance, consider the Amazon-CD dataset: a CD might be categorized under genres like ``folk music'', but it cannot be labeled as ``folk music preferred by teenagers''. However, in reality, such nuanced preferences exist within the collaborative information. When recommending based solely on past interactions, we may overlook these hidden layers of structure that are implicit in the dataset (e.g., ``music'' → ``folk music'' → ``folk music preferred by teenagers'').

To address this, we introduce a hyperbolic hierarchical clustering method based on hyperbolic embeddings to build the hierarchy tree. Specifically, we iteratively group embeddings into clusters and use the centroid of each cluster as the data point to be clustered in the upper layer, referred to as a pseudo-cluster node. However, the number of cluster nodes in each layer is not predefined and must be determined. Inspired by Dasgupta's cost~\cite{dasgupta2016cost}, which encourages pushing edge cuts as far down the tree as possible, the optimal tree is required to be binary. Therefore, we fix the proportion of cluster nodes between layers as $k=2$. Details of the algorithm are presented in Algorithm \ref{alg:cluster}.

\begin{algorithm}
   \caption{Hyperbolic Hierarchical Clustering}
   \label{alg:cluster}
\begin{algorithmic}
    \STATE {\bfseries Input:} Hyperbolic embeddings $X$, layer proportion $k$
    \STATE Set the maximum number of layers $L = \log_k(|X|)$
    \STATE Initialize the points of bottom layer $D_L = X$
    \FOR{$l = L$ {\bfseries to} $1$ (bottom to top)}
        \STATE Randomly initialize cluster centroids in the hyperbolic space $M_l = \{\mu_l\ |\ \mu_l\in \mathbb{H}\}$, $|M_l| = |D_l|/k$
        \REPEAT
            \FOR{each point $d_{l,i} \in D_l$}
                \STATE Assign $d_i$ to the cluster with the nearest centroid:
                \STATE $c_{l,i} = \arg\min_{\mu \in M_l} d_\mathcal{H}(d_i,\mu)$
            \ENDFOR
            \FOR{each cluster $c_{l,i} \in C_l$}
                \STATE Update the centroid $\mu_{l,i}$ as the hyperbolic mean $m_{\mathcal{H}}$ of all points assigned to cluster $c_{l,i}$:
                \STATE $\mu_{l,i} = m_{\mathcal{H}}(\{d\ |\ d\in C_{l,i}\})$
           \ENDFOR
        \UNTIL Centroids $M_l$ do not change (convergence)
        \STATE Initialize the points of the upper layer $D_{l-1} = M_l$
    \ENDFOR
\end{algorithmic}
\end{algorithm}

\subsubsection{Exploration-Exploitation Balance}
To allow users to balance the exploration-exploitation trade-off, we introduce two adjustable parameters: temperature $\tau$ and hierarchy layer $l$. In the standard recommendation approach, items are recommended based on similarity to the user’s embedding, which may limit exploration. For users with focused preferences, this method may consistently recommend items that align too closely with their past choices, offering little opportunity to explore different categories.

To address this issue, we modify the recommendation process by replacing a portion of the recommended items with alternatives from other branches in the hierarchy tree. This modification gives users control over the balance between exploration and exploitation. Temperature $\tau$ allows users to specify the proportion of original recommendations to replace, while layer $l$ determines the starting point for exploration. For example, suppose a user selects $\tau = 0.5$ and $l = 5$. In that case, half of the original recommendations are retained, and the other half is replaced with items sampled from its ancestor cluster at layer $5$ of the hierarchy tree.

\subsection{Complexity Analysis}
\label{sec:com}
In this subsection, we analyze the complexity of constructing and applying the hierarchical representation structure. Let $N$ be the total number of users and items, and $k$ be the number of recommendations per user.

\subsubsection{Hierarchical Clustering}
Building the hierarchy tree involves iterative k-means clustering in hyperbolic space. To optimize efficiency, we assign cluster nodes only to real node embeddings. First, we compute and store pairwise distances, requiring $O(N^2/2)$. Each k-means iteration takes $O(i)$, and with $O(\log N)$ levels, the overall complexity of constructing the hierarchy tree is $O(N^2/2 + i\log N)$.

\subsubsection{Recommendation Retrieval \& Replacement}
Finding the top-$k$ recommendations requires $O(k \log N)$ using nearest neighbour search in hyperbolic space. Adjusting recommendations by retrieving ancestor clusters at layer $l$ requires $O(k)$ time, as it involves only tree traversal.

The proposed hierarchical framework achieves efficient complexity scaling while maintaining flexibility in recommendation adjustments. The hierarchical clustering process, optimized for hyperbolic space, ensures a computationally feasible structure with $O(N^2/2 + i\log N)$ complexity. The recommendation retrieval and replacement steps remain lightweight, running in $O(k \log N)$ and $O(k)$, respectively. This design enables \model to dynamically balance exploration and exploitation while ensuring scalability for large-scale recommender systems.
\section{EXPERIMENTS}
\label{sec:eval}

\begin{table*}[t]
    \centering
    \caption{Comprehensive model comparison based on utility and diversity metrics, where higher values signify superior performance. The best-performing results for each metric are highlighted in bold, with the second-best results underlined.}
    \resizebox{\textwidth}{!}{
        \begin{tabular}{c|c|llll|llllll}
    		\toprule
    		\multirow{2}{*}{Data}& \multirow{2}{*}{Models}& \multicolumn{4}{c|}{Utility $\uparrow$} & \multicolumn{6}{c}{Diveristy $\uparrow$} \\ \cline{3-12} &
            & Recall@10 & NDCG@10 & Recall@20 & NDCG@20 & Div@10 & H@10 & EPC@10 & Div@20 & H@20 & EPC@20 \\ \hline
            \multirow{7}{*}{\shortstack{Amazon-\\books}} &
            \multicolumn{1}{l|}{LightGCN} & 0.0950 & 0.0726 & 0.1428 & 0.0884 & 0.4469 & 9.8554 & 0.7429 & 0.4504 & 10.485 & 0.7798 \\ &
            \multicolumn{1}{l|}{TAG-CF} & 0.0986 & 0.0755 & 0.1468 & 0.0912 & 0.4090 & 10.915 & 0.7899 & 0.4517 & 11.465 & 0.8227 \\ & 
            \multicolumn{1}{l|}{LightGCL} & 0.1075 & 0.0846 & 0.1515 & 0.0991 & 0.4651 & 11.511 & 0.8523 & 0.4974 & 11.941 & 0.8728 \\ &
            \multicolumn{1}{l|}{SimGCL} & 0.0997 & 0.0762 & 0.1478 & 0.0918 & 0.4286 & 10.341 & 0.7757 & 0.4802 & 10.915 & 0.8076 \\ &
            \multicolumn{1}{l|}{HCCF} & 0.1045 & 0.0813 & 0.1488 & 0.0962 & \underline{0.5008} & 11.814 & 0.8579 & \underline{0.5433} & \underline{12.137} & \underline{0.8746} \\ &
            \multicolumn{1}{l|}{HGCF} & 0.1013 & 0.0759 & 0.1511 & 0.0921 & 0.4742 & 11.577 & \underline{0.8623} & 0.5041 & 11.732 & 0.8585 \\ &
            \multicolumn{1}{l|}{HICF} & \underline{0.1113} & \underline{0.0862} & \underline{0.1599} & \underline{0.1022} & 0.4702 & \underline{11.922} & 0.8561 & 0.5261 & 12.129 & 0.8660 \\ &
            \multicolumn{1}{l|}{Ours} & \textbf{0.1167} & \textbf{0.0902} & \textbf{0.1682} & \textbf{0.1069} &\textbf{0.5255} & \textbf{12.096} & \textbf{0.8700} & \textbf{0.5754} & \textbf{12.254} & \textbf{0.8758} \\
            \hline
            \multirow{7}{*}{Yelp} &
            \multicolumn{1}{l|}{LightGCN} & 0.0678 & 0.0565 & 0.1135 & 0.0717 & 0.2603 & 8.3263 & 0.6842 & 0.2277 & 9.1811 & 0.7343 \\ &
            \multicolumn{1}{l|}{TAG-CF} & 0.0692 & 0.0571 & 0.1142 & 0.0722 & 0.2353 & 8.4516 & 0.6939 & 0.2276 & 9.3108 & 0.7428 \\ & 
            \multicolumn{1}{l|}{LightGCL} & \underline{0.0709} & \underline{0.0594} & 0.1141 & 0.0739 & 0.2856 & 11.243 & 0.8281 & 0.3064 & 11.823 & 0.8550 \\ &
            \multicolumn{1}{l|}{SimGCL} & 0.0707 & 0.0586 & \underline{0.1184} & \underline{0.0748} & 0.3397 & 9.8191 & 0.7642 & 0.3460 & 10.508 & 0.7973 \\ &
            \multicolumn{1}{l|}{HCCF} & 0.0680 & 0.0562 & 0.1097 & 0.0701 & 0.3168 & 11.942 & 0.8572 & 0.3535 & \underline{12.311} & \underline{0.8752} \\ &
            \multicolumn{1}{l|}{HGCF} & 0.0654 & 0.0510 & 0.1099 & 0.0665 & 0.3133 & \underline{11.977} & \textbf{0.8718} & 0.3289 & 12.103 & 0.8747 \\ &
            \multicolumn{1}{l|}{HICF} & \underline{0.0709} & 0.0579 & 0.1169 & 0.0737 & \underline{0.3478} & 11.877 & 0.8557 & \underline{0.3721} & 12.147 & 0.8666 \\ &
            \multicolumn{1}{l|}{Ours} & \textbf{0.0739} & \textbf{0.0597} & \textbf{0.1224} & \textbf{0.0762} & \textbf{0.3874} & \textbf{12.155} & \underline{0.8716} & \textbf{0.3961} & \textbf{12.329} & \textbf{0.8773} \\
            \hline
            \multirow{7}{*}{\shortstack{Google-\\reviews}} &
            \multicolumn{1}{l|}{LightGCN} & 0.0932 & 0.0745 & 0.1430 & 0.0929 & 0.2422 & 9.9149 & 0.7241 & 0.2725 & 10.761 & 0.7749 \\ &
            \multicolumn{1}{l|}{TAG-CF} & 0.0942 & 0.0754 & 0.1447 & 0.0940 & 0.2376 & 10.184 & 0.7485 & 0.2497 & 11.031 & 0.7961 \\ & 
            \multicolumn{1}{l|}{LightGCL} & 0.0945 & 0.0768 & 0.1461 & 0.0959 & 0.2639 & 11.643 & 0.8333 & 0.2705 & 12.317 & \underline{0.8655} \\ &
            \multicolumn{1}{l|}{SimGCL} & \underline{0.0979} & \textbf{0.0784} & 0.1482 & \underline{0.0971} & 0.2966 & 10.361 & 0.7486 & 0.2950 & 11.119 & 0.7923 \\ &
            \multicolumn{1}{l|}{HCCF} & 0.0923 & 0.0744 & 0.1446 & 0.0937 & 0.2195 & 10.741 & 0.8056 & 0.2402 & 11.561 & 0.8433 \\ &
            \multicolumn{1}{l|}{HGCF} & 0.0910 & 0.0697 & 0.1485 & 0.0909 & 0.2859 & 12.248 & \underline{0.8538} & 0.3321 & 12.474 & 0.8619 \\ &
            \multicolumn{1}{l|}{HICF} & 0.0951 & 0.0736 & \underline{0.1515} & 0.0944 & \textbf{0.3262} & \underline{12.376} & 0.8523 & \underline{0.3510} & \underline{12.660} & 0.8640 \\ &
            \multicolumn{1}{l|}{Ours} & \textbf{0.1002} & \underline{0.0772} & \textbf{0.1598} & \textbf{0.0992} & \underline{0.3185} & \textbf{12.466} & \textbf{0.8598} & \textbf{0.3543} & \textbf{12.769} & \textbf{0.8728} \\
            \bottomrule
        \end{tabular}
    }
    \label{tab:General Performance}
\end{table*}

\subsection{Experimental Settings}
\subsubsection{Datasets}
To evaluate our proposed model, we utilize three widely-used public datasets: \textbf{Amazon-books}, which contains user purchase behaviors within the book category on Amazon; \textbf{Yelp}, which records customer ratings and reviews for restaurants on Yelp; and \textbf{Google-reviews}, which includes user reviews and business metadata from Google Maps. The high proportion of tail users (T80) highlights the prevalence of long-tail distributions, supporting the assumption of a power-law distribution in real-world data. Detailed statistics, including the distribution of head and tail items, are provided in Table ~\ref{tab:datasets}.

\begin{table}[t]
    \caption{Statistics of the experimental data.}
    \label{tab:datasets}
    \centering
    \resizebox{\linewidth}{!}{
    \begin{tabular}{@{}ccccccc@{}}
        \toprule
        \multirow{2}{*}{Dataset} & \multirow{2}{*}{\#User} & \multicolumn{3}{c}{\#Item} & \multirow{2}{*}{\#Interactions} & \multirow{2}{*}{Density} \\ \cmidrule(lr){3-5} & & All & H20(\%) & T80(\%) & & \\ \midrule
        Amazon-books & 11,000 & 9,332 & 45.2 & 54.8 & 120,464 & 1.17$e^{-3}$ \\
        Yelp & 11,091 & 11,010 & 47.0 & 53.0 & 166,620 & 1.36$e^{-3}$ \\
        Google-reviews & 22,582 & 16,557 & 46.8 & 53.2 & 411,840 & 1.10$e^{-3}$ \\
        \bottomrule
    \end{tabular}%
    }
    \label{tab:statistics}
\end{table}

To further evaluate the scalability and practical utility of our proposed method in real-world deployment scenarios, we conduct additional experiments on two substantially larger datasets: \textbf{Amazon-Books (Large)} and \textbf{Amazon-CDs (Large)}. These datasets represent a significant increase in scale, containing up to ten times the number of users and interactions compared to the standard benchmarks utilized in previous work. Comprehensive statistical details and the corresponding experimental results are provided in Section \ref{subsec: larger dataset}.

\subsubsection{Evaluation Metrics}
We assess our recommended items using both utility and diversity metrics, demonstrating that our model achieves precise recommendations for users' preference while enhancing diversity in the recommendation. For utility, we adopt two widely used ranking metrics: Recall and NDCG, which measure the effectiveness of the model. To evaluate diversity, we employ three metrics: Distance Diversity (Div)~\cite{ziegler2005improving, zhang2008avoiding}, Shannon Entropy (H)~\cite{mehta2012collaborative}, and Expected Popularity Complement (EPC)~\cite{vargas2011rank}.

Distance Diversity (Div) quantifies the variation among recommended items, encouraging diversity within the recommendation:
\begin{equation}
    Div(u) = \frac{\sum_{i \in R_u} \sum_{j \in R_u, j \neq i} d(i, j)}{\text{Number of } (i, j) \text{ pairs}}
\end{equation}
where $R_u$ is the recommendation list for user $u$, and $d(i, j)$ represents the distance between items $i$ and $j$. Number of $(i, j)$ pairs is calculated as $\binom{|R_u|}{2} = \frac{|R_u|(|R_u|-1)}{2}$. For fair comparison, we compute distances within the same representation space across models.

Shannon Entropy (H) quantifies the unpredictability within the recommendations, with higher entropy signifying greater diversity:
\begin{equation}
    H = -\sum_{i \in \text{set}(\sum_u R_u)} p(i) \log p(i)
\end{equation}
where $\sum_u R_u$ denotes all recommended items across users (including duplicates), and $p(i) = \text{count}(i)/\sum_{i \in \text{set}(\sum_u R_u)} \text{count}(i)$ is the proportion of item $i$ among all recommended items.

Expected Popularity Complement (EPC) assesses the extent of bias towards popular items by promoting niche recommendations:
\begin{equation}
    EPC = 1 - \frac{1}{|\text{set}(\sum R_u)|} \sum_{i \in \text{set}(\sum_u R_u)} \frac{\text{pop}(i)}{\max_i(\text{pop}(i))}
\end{equation}
where $N_u$ is the set of items user $u$ has interacted with, and $\text{pop}(i) = \text{count}(i)/\sum_{i \in \text{set}(\sum_u N_u)} \text{count}(i)$ denotes item $i$'s popularity.

\subsubsection{Baselines}
We compare our model with state-of-the-art Euclidean and Hyperbolic baselines:
\begin{itemize}[leftmargin=*, itemsep=0pt]
    \item \textbf{LightGCN}~\cite{he2020lightgcn}: A lightweight graph convolution neural network model for recommendation by removing redundant neural modules in graph convolution.
    \item \textbf{TAG-CF}~\cite{ju2024does}: A test-time augmentation framework that enhances recommendation by performing a single message-passing step only at inference time.
    \item \textbf{LightGCL}~\cite{cai2023lightgcl}: A contrastive learning model that uses singular value decomposition (SVD) for robust graph augmentation without data augmentation.
    \item \textbf{SimGCL}~\cite{yu2022graph}: A contrastive learning model that replaces graph augmentations in contrastive learning by introducing random noise during initialization.
    \item \textbf{HCCF}~\cite{xia2022hypergraph}: A collaborative filtering framework that integrates hypergraph and contrastive learning, capturing both local and global user dependencies.
    \item \textbf{HGCF}~\cite{sun2021hgcf}: A hyperbolic GNN model that combines hypergraph learning with collaborative filtering using hyperbolic margin ranking loss.
    \item \textbf{HICF}~\cite{yang2022hicf}: A hyperbolic recommendation model that enhances head and tail item performance with hyperbolic geometry, improving utility for long-tail items.
\end{itemize}

\subsubsection{Implementation Details}
\label{subsec:implementation details}
We set the embedding dimension to 50 and the training batch size to 1024. Hyperbolic space embeddings are optimized using Riemannian SGD (RSGD)~\cite{becigneul2018riemannian, sun2021hgcf}, while the MLP adapter is trained via Adam~\cite{kingma2014adam}. Training employs early stopping based on Recall@10 with a patience of 10 epochs. For the Amazon-Books, Amazon-Books (Large), Amazon-CDs (Large), and Yelp datasets, user and item profiles are generated via GPT-3.5-turbo and encoded using text-embedding-ada-002~\cite{neelakantan2022text}. To verify encoder robustness, we substitute the text encoder with BERT~\cite{devlin2018bert} for Google-Reviews, where our method consistently maintains its performance superiority. 

All experiments are executed on NVIDIA A40/A100 GPUs (40GB VRAM). The framework demonstrates strong computational efficiency: training requires fewer than 5 seconds per epoch on standard benchmarks (Amazon-Books, Yelp, Google-Reviews), and scales efficiently to larger corpora, requiring approximately 45 seconds for Amazon-Books (Large) and 20 seconds for Amazon-CDs (Large) per epoch.

For final evaluations, we conduct a grid search over key hyperparameters: learning rate $\in \{1\mathrm{e}{-3}, 5\mathrm{e}{-3}, 1\mathrm{e}{-2}, 5\mathrm{e}{-2}, 1\mathrm{e}{-1}\}$, weight decay $\in \{1\mathrm{e}{-3}, 1\mathrm{e}{-2}, 1\mathrm{e}{-1}\}$, number of negative samples $\in \{10, 20, 50\}$, and alignment weight $\in \{1\mathrm{e}{-3}, 1\mathrm{e}{-2}, 1\mathrm{e}{-1}\}$. The complete hyperparameter logs are available in our attached repository.

\subsection{Performance Comparison}
\subsubsection{Overall Comparison}  
The performance comparison across all models is presented in Table \ref{tab:General Performance}, with statistical significance testing in Appendix \ref{subsec: Statistical Significance Testing}. Our \model\ consistently outperforms the baselines in both utility and diversity metrics, achieving up to $5.49\%$ improvement in utility metrics and $11.39\%$ increase in diversity metrics. Notably, most baselines struggle to achieve strong performance in both utility and diversity simultaneously. For instance, while HICF performs second-best in utility metrics on the Amazon-books dataset, it lacks the same advantage in diversity metrics, where several baselines achieve better results. This highlights the superiority of our model, which achieves state-of-the-art performance in both utility and diversity: a feat not previously attained by any baseline.

\begin{table*}[t]
    \centering
    \caption{Comparison with feature-enhanced Euclidean baselines, with the best-performing results highlighted in bold.}
    \resizebox{\textwidth}{!}{
        \begin{tabular}{l|lllll|lllll|lllll}
            \toprule
                \multicolumn{1}{l|}{Data} & \multicolumn{5}{c|}{Amazon-books} & \multicolumn{5}{c|}{Yelp} & \multicolumn{5}{c}{Google-reviews} \\
            \hline
                \multicolumn{1}{l|}{Metrics} & R@20 & N@20 & Div@20 & H@20 & EPC@20 & R@20 & N@20 & Div@20 & H@20 & EPC@20 & R@20 & N@20 & Div@20 & H@20 & EPC@20 \\
                \hline
                LightGCN+ & 0.1466 & 0.0901 & 0.4349 & 10.3915 & 0.7802 & 0.1156 & 0.0733 & 0.2697 & 9.6903 & 0.7526 & 0.1438 & 0.0940 & 0.2707 & 10.8664 & 0.7801 \\
                LightGCL+ & 0.1545 & 0.0992 & 0.5064 & 11.9761 & 0.8728 & 0.1131 & 0.0739 & 0.2759 & 11.8617 & 0.8581 & 0.1443 & 0.0946 & 0.2744 & 12.4700 & \textbf{0.8776} \\
                SimGCL+ & 0.1494 & 0.0926 & 0.5163 & 11.4751 & 0.8342 & 0.1195 & 0.0752 & 0.3152 & 9.9952 & 0.7811 & 0.1504 & 0.0982 & 0.3030 & 11.2111 & 0.7953 \\
                HCCF+ & 0.1525 & 0.0940 & 0.4316 & 11.7984 & 0.8616 & 0.1170 & 0.0729 & 0.2884 & 11.6906 & 0.8577 & 0.1438 & 0.0941 & 0.2524 & 11.5433 & 0.8406 \\
                Ours & \textbf{0.1682} & \textbf{0.1069} & \textbf{0.5754} & \textbf{12.2543} & \textbf{0.8758} & \textbf{0.1224} & \textbf{0.0762} & \textbf{0.3961} & \textbf{12.3290} & \textbf{0.8773} & \textbf{0.1598} & \textbf{0.0992} & \textbf{0.3543} & \textbf{12.7690} & 0.8728 \\
            \bottomrule
        \end{tabular}
    }
    \label{tab:fair comparison}
\end{table*}

\subsubsection{Baselines with Feature Enhancement}  
To ensure a fair comparison, we further enhance Euclidean baselines by integrating feature information. Specifically, we align semantic embeddings with the original Euclidean embeddings following the structure in Section~\ref{subsec: Semantic Representations Generation}, but in Euclidean space. The results are presented in Table \ref{tab:fair comparison}, with enhanced baselines denoted by ``+''. The data shows that our \model\ still surpasses all baselines in both utility and diversity. Interestingly, while the feature-enhanced versions of the Euclidean baselines generally improve in utility metrics, they do not exhibit a stable improvement in diversity metrics. This aligns with our theoretical analysis in Section~\ref{sub:gradient analysis}, which demonstrates the advantages of our alignment in the hyperbolic space.


\begin{figure}[t]
    \centering
    \includegraphics[width=\columnwidth]{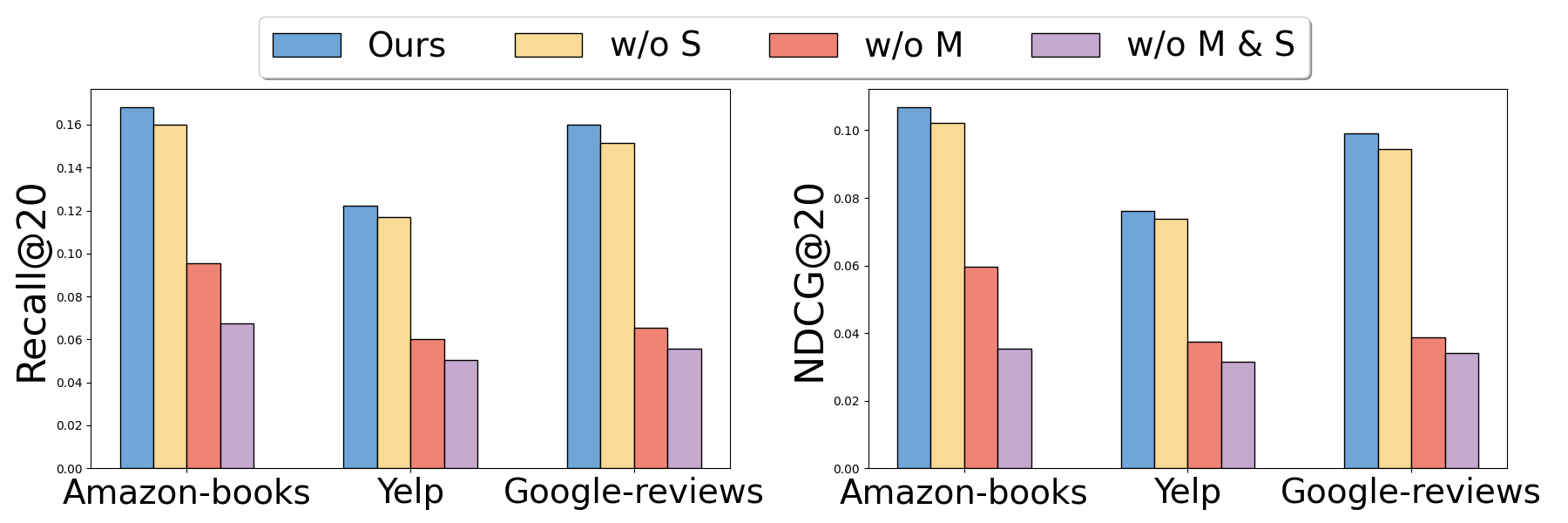}
    \caption{Ablation study on model variants.}
    \label{fig:ablation}
    \vskip -0.1in
\end{figure}

\subsubsection{Performance on Head and Tail Items}
To underscore the effectiveness of semantic alignment, particularly for cold-start (tail) items, we conducted experiments measuring utility performance on both head and tail items. In our context, head items (H20) refer to the top $20\%$ of items with the highest number of interactions in the training dataset, representing the most popular items. The remaining items are categorized as tail items (T80), which have limited interaction history, representing less frequently engaged items.

We compare our model with all hyperbolic baselines, as well as a representative Euclidean baseline. Since Euclidean models perform significantly worse on tail items, our primary focus is on comparisons within hyperbolic space. The results, displayed in Table \ref{tab:Head and Tail}, reveal significant improvements, especially for tail items. Compared with hyperbolic baselines without semantic alignment, our model shows a notably larger performance boost on tail items than on head items. This finding indicates that semantic alignment particularly enhances the recommendation of cold-start items, effectively mitigating the cold-start problem for items with limited or no prior interactions.

Specifically, our model achieves substantial improvements on tail items, with utility gains reaching up to $9.71\%$ on Amazon-Books, $13.33\%$ on Yelp, and $14.42\%$ on Google-Reviews. These results highlight that semantic alignment not only improves overall recommendation utility but also significantly enhances the model’s ability to recommend niche or less popular items, creating a balanced recommendation list that includes both popular and cold-start items. This capability is critical in real-world recommendation settings, where new or less-interacted items benefit from enhanced visibility and user interaction, ultimately enriching the user experience with a broader and more diverse range of options.

\subsection{Ablation Study}
We perform an ablation study to examine the contributions of two key components in our model: hyperbolic margin ranking loss and semantic alignment loss. To maintain meaningful evaluation, we focus solely on utility metrics, as random recommendations may artificially inflate diversity without practical utility. The following notations are used for our ablation variants: ``w/o M'' denotes the model without hyperbolic margin ranking loss, ``w/o S'' represents the model without semantic alignment loss, and ``w/o M \& S'' indicates that both margin ranking and semantic alignment are excluded, leaving only the BPR loss. Figure \ref{fig:ablation} shows that our full model outperforms all ablation variants, highlighting its overall effectiveness. Models lacking both margin ranking and semantic alignment perform the worst, while removing either module results in decreased utility, emphasizing the effectiveness of integration.

Recent work \cite{yang2024hgformer, yang2024hyperbolic} has proposed performing message passing directly in hyperbolic space as an alternative to using the tangent space. We implemented this approach in place of our original hyperbolic graph collaborative filtering method. As shown in Appendix \ref{subsec: Alternative hyperbolic message passing}, this alternative achieves comparable performance, making it a viable alternative.

\begin{figure}[t]
    \centering
    \includegraphics[width=\columnwidth]{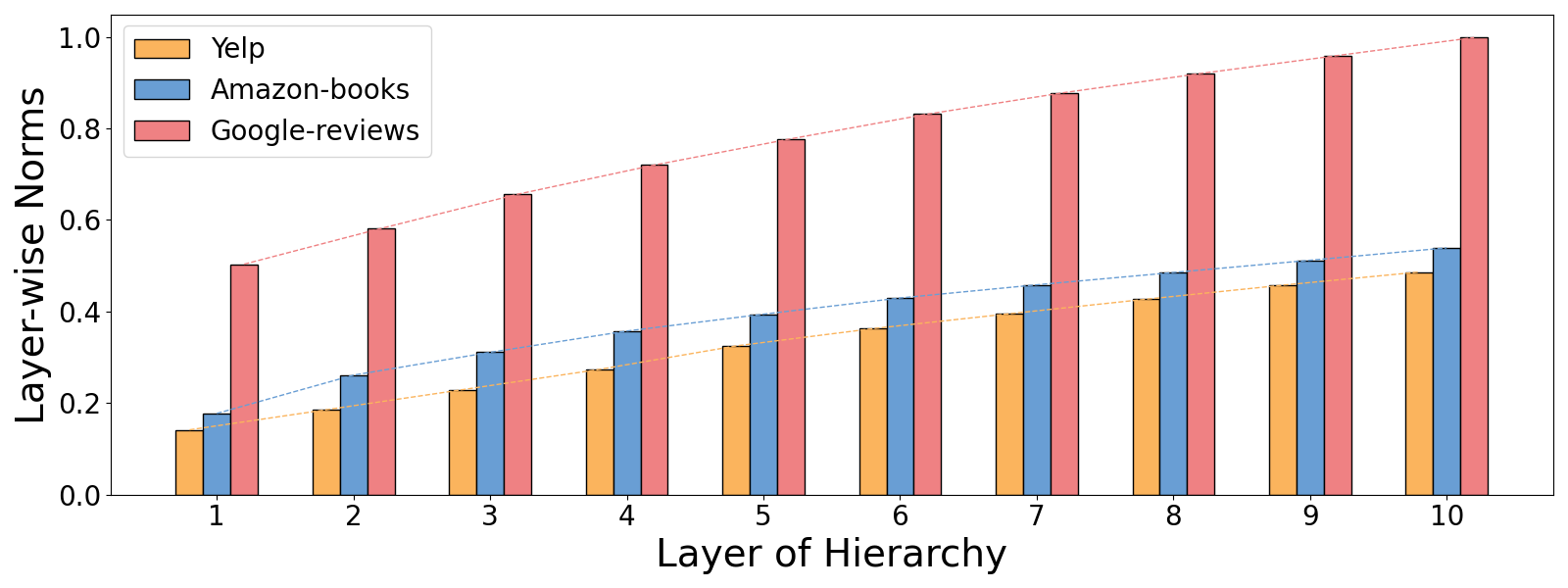}
    \vskip -0.1in
    \caption{Layer-wise norms.}
    \label{fig:norm}
    \vskip -0.1in
\end{figure}

\begin{table*}[t]
    \centering
    \caption{Performance on H20 (head) and T80 (tail) items, with the top hyperbolic results highlighted in bold. $\Delta_\mathcal{H}(\%)$ represents the improvement over the second-best hyperbolic baseline (underlined).}
    \begin{tabular}{c|ll|ll|ll|ll|ll|ll}
        \toprule
        Datasets & \multicolumn{4}{c|}{Amazon-books} & \multicolumn{4}{c|}{Yelp} & \multicolumn{4}{c}{Google-reviews}\\ \cline{1-13}
        \multirow{2}{*}{Metrics} & \multicolumn{2}{c|}{Recall@20} & \multicolumn{2}{c|}{NDCG@20} & \multicolumn{2}{c|}{Recall@20} & \multicolumn{2}{c|}{NDCG@20} & \multicolumn{2}{c|}{Recall@20} & \multicolumn{2}{c}{NDCG@20} \\
        & H20 & T80 & H20 & T80 & H20 & T80 & H20 & T80 & H20 & T80 & H20 & T80 \\ \hline
        \multicolumn{1}{c|}{LightGCN} & 0.1162 & 0.0266 & 0.0744 & 0.0140 & 0.1076 & 0.0059 & 0.0693 & 0.0024 & 0.1311 & 0.0119 & 0.0878 & 0.0051 \\ 
        \multicolumn{1}{l|}{HGCF} & 0.1083 & 0.0428 & 0.0673 & 0.0248 & 0.0789 & 0.0310 & 0.0489 & 0.0176 & 0.1123 & 0.0362 & 0.0714 & 0.0195 \\
        \multicolumn{1}{l|}{HICF} & \underline{0.1086} & \underline{0.0513} & \underline{0.0713} & \underline{0.0309} & \underline{0.0847} & \underline{0.0322} & \underline{0.0556} & \underline{0.0180} & \underline{0.1147} & \underline{0.0388} & \underline{0.0746} & \underline{0.0208} \\
        \multicolumn{1}{l|}{Ours} & \textbf{0.1125} & \textbf{0.0557} & \textbf{0.0730} & \textbf{0.0339} & \textbf{0.0863} & \textbf{0.0361} & \textbf{0.0558} & \textbf{0.0204} & \textbf{0.1155} & \textbf{0.0443} & \textbf{0.0755} & \textbf{0.0238} \\
        \multicolumn{1}{l|}{$\Delta_\mathcal{H}(\%)$} & +3.59 & +8.58 & +2.38 & +9.71 & +1.89 & +12.11 & +0.36 & +13.33 & +0.70 & +14.18 & +1.21 & +14.42 \\
        \bottomrule
    \end{tabular}
    \label{tab:Head and Tail}
\end{table*}

\begin{figure*}[t]
    \centering
    \includegraphics[width=\textwidth]{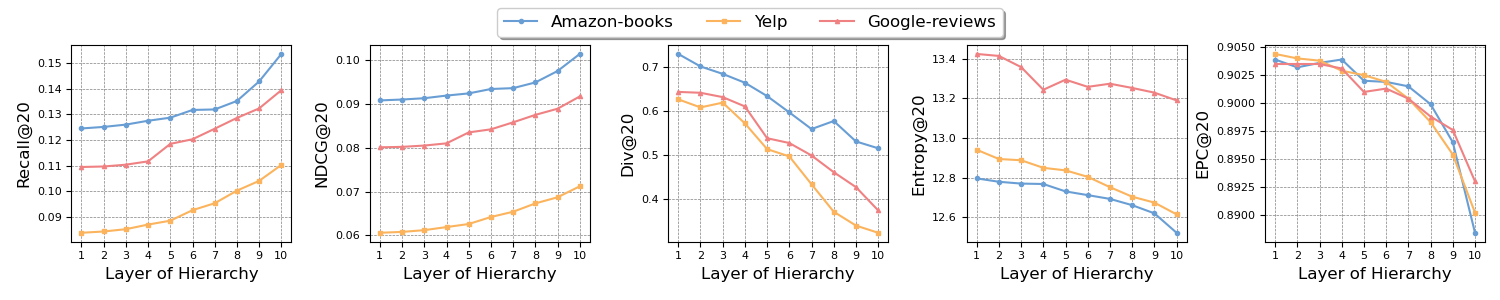}
    \caption{Performance analysis across hierarchical structure layers.}
    \label{fig:explore}
\end{figure*}

\subsection{Hierarchical Structure Analysis}
We evaluated the effectiveness of our hierarchical representation structure through experiments simulating varying user preferences. Using a temperature parameter $\tau = 0.5$ to control recommendation randomness, we analyzed performance across various hierarchical layers. As shown in Figure \ref{fig:explore}, utility metrics increase while diversity metrics decrease with larger layers. This trade-off aligns with the model's design: larger layers focus on utility and specific interests for exploitation, while smaller layers prioritize diversity for exploration. This structure provides users with a flexible mechanism to adjust recommendations from niche, relevant items to broader, exploratory sets based on their preferences.

To further validate the hierarchical structure's capability to capture varying degrees of group preferences, we analyzed the average norm of node embeddings across layers (Figure \ref{fig:norm}). The results reveal a consistent increase at larger layers (farther from the center in hyperbolic space). As previously mentioned, hyperbolic geometry ensures that popular items or exploratory users are positioned closer to the origin, while niche items or users are positioned toward the boundaries. This observation aligns with the theoretical properties of hyperbolic space and supports our hierarchical design: higher-level (smaller-layer) nodes effectively encapsulate abstract and aggregated preferences, whereas lower-level (larger-layer) nodes focus on fine-grained, personalized preferences. 

Since quantitatively evaluating the performance of unsupervised clustering remains challenging, we further present a case study below to demonstrate the effectiveness of our approach.


\subsection{Hierarchical Structure Case Study}
\label{sec:case study}
In this subsection, we present a case study examining the exploration-exploitation trade-off in the hyperbolic hierarchical structure. We randomly select a user and identify items that share different layers of the lowest common ancestor (LCA) with the user in the hierarchy tree. Ideally, a larger LCA layer indicates that the user and the item are grouped earlier in the hierarchy, suggesting a stronger likelihood of interaction. Conversely, a smaller LCA layer implies that the user and the item can only be grouped after many iterations, indicating a weaker likelihood of interaction.\\
\hrule
~\\
\noindent \textit{\textbf{User Profile:}\\
The user enjoys books with supernatural elements, horror and paranormal romance. They also appreciate fast-paced action and suspenseful thrillers with political intrigue and espionage. The user likes complex characters dealing with dark themes.\\~\\
\textbf{Item 1 Profile (LCA=10):}\\
The book 'Spirit Fighter (Son of Angels, Jonah Stone)' would likely appeal to young readers who enjoy stories about supernatural beings and the fight between good and evil. It is a book that offers entertainment and inspiration, as well as a glimpse into the supernatural world.\\~\\
\textbf{Item 2 Profile (LCA=5):}\\
Fans of contemporary romance novels will enjoy Barefoot Summer (A Chapel Springs Romance). This book combines heartwarming romance, relatable heartaches, and lighthearted humor to create a delightful and engaging story. It is ideal for those looking for an easy-to-read and entertaining summer novel.}\\
\hrule
~\\

As demonstrated in the case study, the user shows a preference for supernatural elements, which aligns closely with the description of item 1, indicating a high likelihood of interaction. In contrast, while the user profile mentions a preference for dark themes, item 2 focuses on romance and heartwarming narratives, suggesting a low probability of user interest in item 2.

\section{CONCLUSION}
\label{sec:conclusion}

In this work, we present \model, a hyperbolic framework that effectively balances exploration and exploitation in recommender systems. Our approach integrates semantic and collaborative information within hyperbolic space, leveraging the unique properties of hyperbolic geometry, supported by rigorous mathematical proof. Furthermore, we introduce a personalized mechanism for balancing the exploration-exploitation trade-off using a hierarchical representation structure, which is a hyperparameter-free clustering mechanism theoretically optimized by Dasgupta’s cost, empowering users to adjust recommendation preferences flexibly. This functionality is particularly valuable in real-world applications where user preferences may shift dynamically, and our model’s adaptability to these changes contributes to a more personalized recommendation experience. Extensive experiments demonstrate that \model achieves state-of-the-art performance, delivering up to a $5.49\%$ improvement in utility metrics and an $11.39\%$ increase in diversity metrics, effectively mitigating information cocoons.

\clearpage
\bibliographystyle{plainnat}
\bibliography{refs}

@inproceedings{he2020lightgcn,
  title={Lightgcn: Simplifying and powering graph convolution network for recommendation},
  author={He, Xiangnan and Deng, Kuan and Wang, Xiang and Li, Yan and Zhang, Yongdong and Wang, Meng},
  booktitle={Proceedings of the 43rd International ACM SIGIR conference on research and development in Information Retrieval},
  pages={639--648},
  year={2020}
}

@inproceedings{yang2022hicf,
  title={Hicf: Hyperbolic informative collaborative filtering},
  author={Yang, Menglin and Li, Zhihao and Zhou, Min and Liu, Jiahong and King, Irwin},
  booktitle={Proceedings of the 28th ACM SIGKDD Conference on Knowledge Discovery and Data Mining},
  pages={2212--2221},
  year={2022}
}

@article{krioukov2010hyperbolic,
  title={Hyperbolic geometry of complex networks},
  author={Krioukov, Dmitri and Papadopoulos, Fragkiskos and Kitsak, Maksim and Vahdat, Amin and Bogun{\'a}, Mari{\'a}n},
  journal={Physical Review E—Statistical, Nonlinear, and Soft Matter Physics},
  volume={82},
  number={3},
  pages={036106},
  year={2010},
  publisher={APS}
}

@inproceedings{sun2021hgcf,
  title={Hgcf: Hyperbolic graph convolution networks for collaborative filtering},
  author={Sun, Jianing and Cheng, Zhaoyue and Zuberi, Saba and P{\'e}rez, Felipe and Volkovs, Maksims},
  booktitle={Proceedings of the Web Conference 2021},
  pages={593--601},
  year={2021}
}

@article{ma2024xrec,
  title={XRec: Large Language Models for Explainable Recommendation},
  author={Ma, Qiyao and Ren, Xubin and Huang, Chao},
  journal={arXiv preprint arXiv:2406.02377},
  year={2024}
}

@inproceedings{ren2024representation,
  title={Representation learning with large language models for recommendation},
  author={Ren, Xubin and Wei, Wei and Xia, Lianghao and Su, Lixin and Cheng, Suqi and Wang, Junfeng and Yin, Dawei and Huang, Chao},
  booktitle={Proceedings of the ACM on Web Conference 2024},
  pages={3464--3475},
  year={2024}
}

@article{becigneul2018riemannian,
  title={Riemannian adaptive optimization methods},
  author={B{\'e}cigneul, Gary and Ganea, Octavian-Eugen},
  journal={arXiv preprint arXiv:1810.00760},
  year={2018}
}

@inproceedings{xia2022hypergraph,
  title={Hypergraph contrastive collaborative filtering},
  author={Xia, Lianghao and Huang, Chao and Xu, Yong and Zhao, Jiashu and Yin, Dawei and Huang, Jimmy},
  booktitle={Proceedings of the 45th International ACM SIGIR conference on research and development in information retrieval},
  pages={70--79},
  year={2022}
}

@inproceedings{yu2022graph,
  title={Are graph augmentations necessary? simple graph contrastive learning for recommendation},
  author={Yu, Junliang and Yin, Hongzhi and Xia, Xin and Chen, Tong and Cui, Lizhen and Nguyen, Quoc Viet Hung},
  booktitle={Proceedings of the 45th international ACM SIGIR conference on research and development in information retrieval},
  pages={1294--1303},
  year={2022}
}

@article{cai2023lightgcl,
  title={LightGCL: Simple yet effective graph contrastive learning for recommendation},
  author={Cai, Xuheng and Huang, Chao and Xia, Lianghao and Ren, Xubin},
  journal={arXiv preprint arXiv:2302.08191},
  year={2023}
}

@inproceedings{ziegler2005improving,
  title={Improving recommendation lists through topic diversification},
  author={Ziegler, Cai-Nicolas and McNee, Sean M and Konstan, Joseph A and Lausen, Georg},
  booktitle={Proceedings of the 14th international conference on World Wide Web},
  pages={22--32},
  year={2005}
}

@inproceedings{zhang2008avoiding,
  title={Avoiding monotony: improving the diversity of recommendation lists},
  author={Zhang, Mi and Hurley, Neil},
  booktitle={Proceedings of the 2008 ACM conference on Recommender systems},
  pages={123--130},
  year={2008}
}

@inproceedings{vargas2011rank,
  title={Rank and relevance in novelty and diversity metrics for recommender systems},
  author={Vargas, Sa{\'u}l and Castells, Pablo},
  booktitle={Proceedings of the fifth ACM conference on Recommender systems},
  pages={109--116},
  year={2011}
}

@article{mehta2012collaborative,
  title={Collaborative personalized web recommender system using entropy based similarity measure},
  author={Mehta, Harita and Bhatia, Shveta Kundra and Bedi, Punam and Dixit, Veer Sain},
  journal={arXiv preprint arXiv:1201.4210},
  year={2012}
}

@article{rendle2012bpr,
  title={BPR: Bayesian personalized ranking from implicit feedback},
  author={Rendle, Steffen and Freudenthaler, Christoph and Gantner, Zeno and Schmidt-Thieme, Lars},
  journal={arXiv preprint arXiv:1205.2618},
  year={2012}
}

@inproceedings{wang2019neural,
  title={Neural graph collaborative filtering},
  author={Wang, Xiang and He, Xiangnan and Wang, Meng and Feng, Fuli and Chua, Tat-Seng},
  booktitle={Proceedings of the 42nd international ACM SIGIR conference on Research and development in Information Retrieval},
  pages={165--174},
  year={2019}
}

@inproceedings{wu2021self,
  title={Self-supervised graph learning for recommendation},
  author={Wu, Jiancan and Wang, Xiang and Feng, Fuli and He, Xiangnan and Chen, Liang and Lian, Jianxun and Xie, Xing},
  booktitle={Proceedings of the 44th international ACM SIGIR conference on research and development in information retrieval},
  pages={726--735},
  year={2021}
}

@article{koren2009matrix,
  title={Matrix factorization techniques for recommender systems},
  author={Koren, Yehuda and Bell, Robert and Volinsky, Chris},
  journal={Computer},
  volume={42},
  number={8},
  pages={30--37},
  year={2009},
  publisher={IEEE}
}

@inproceedings{yang2022hrcf,
  title={HRCF: Enhancing collaborative filtering via hyperbolic geometric regularization},
  author={Yang, Menglin and Zhou, Min and Liu, Jiahong and Lian, Defu and King, Irwin},
  booktitle={Proceedings of the ACM Web Conference 2022},
  pages={2462--2471},
  year={2022}
}

@article{liu2019hyperbolic,
  title={Hyperbolic graph neural networks},
  author={Liu, Qi and Nickel, Maximilian and Kiela, Douwe},
  journal={Advances in neural information processing systems},
  volume={32},
  year={2019}
}

@article{chami2019hyperbolic,
  title={Hyperbolic graph convolutional neural networks},
  author={Chami, Ines and Ying, Zhitao and R{\'e}, Christopher and Leskovec, Jure},
  journal={Advances in neural information processing systems},
  volume={32},
  year={2019}
}

@inproceedings{ying2018graph,
  title={Graph convolutional neural networks for web-scale recommender systems},
  author={Ying, Rex and He, Ruining and Chen, Kaifeng and Eksombatchai, Pong and Hamilton, William L and Leskovec, Jure},
  booktitle={Proceedings of the 24th ACM SIGKDD international conference on knowledge discovery \& data mining},
  pages={974--983},
  year={2018}
}

@inproceedings{zhang2021lorentzian,
  title={Lorentzian graph convolutional networks},
  author={Zhang, Yiding and Wang, Xiao and Shi, Chuan and Liu, Nian and Song, Guojie},
  booktitle={Proceedings of the web conference 2021},
  pages={1249--1261},
  year={2021}
}

@inproceedings{du2022hakg,
  title={HAKG: Hierarchy-aware knowledge gated network for recommendation},
  author={Du, Yuntao and Zhu, Xinjun and Chen, Lu and Zheng, Baihua and Gao, Yunjun},
  booktitle={Proceedings of the 45th international ACM SIGIR conference on Research and development in Information Retrieval},
  pages={1390--1400},
  year={2022}
}

@inproceedings{tai2021knowledge,
  title={Knowledge based hyperbolic propagation},
  author={Tai, Chang-You and Huang, Chien-Kun and Huang, Liang-Ying and Ku, Lun-Wei},
  booktitle={Proceedings of the 44th International ACM SIGIR Conference on Research and Development in Information Retrieval},
  pages={1945--1949},
  year={2021}
}

@inproceedings{chen2021neural,
  title={Neural collaborative reasoning},
  author={Chen, Hanxiong and Shi, Shaoyun and Li, Yunqi and Zhang, Yongfeng},
  booktitle={Proceedings of the Web Conference 2021},
  pages={1516--1527},
  year={2021}
}

@article{ju2024does,
  title={How Does Message Passing Improve Collaborative Filtering?},
  author={Ju, Mingxuan and Shiao, William and Guo, Zhichun and Ye, Yanfang and Liu, Yozen and Shah, Neil and Zhao, Tong},
  journal={arXiv preprint arXiv:2404.08660},
  year={2024}
}

@article{devlin2018bert,
  title={Bert: Pre-training of deep bidirectional transformers for language understanding},
  author={Devlin, Jacob},
  journal={arXiv preprint arXiv:1810.04805},
  year={2018}
}

@article{wilson2021balancing,
  title={Balancing exploration and exploitation with information and randomization},
  author={Wilson, Robert C and Bonawitz, Elizabeth and Costa, Vincent D and Ebitz, R Becket},
  journal={Current opinion in behavioral sciences},
  volume={38},
  pages={49--56},
  year={2021},
  publisher={Elsevier}
}

@article{lyu2023llm,
  title={Llm-rec: Personalized recommendation via prompting large language models},
  author={Lyu, Hanjia and Jiang, Song and Zeng, Hanqing and Xia, Yinglong and Wang, Qifan and Zhang, Si and Chen, Ren and Leung, Christopher and Tang, Jiajie and Luo, Jiebo},
  journal={arXiv preprint arXiv:2307.15780},
  year={2023}
}

@inproceedings{bao2023tallrec,
  title={Tallrec: An effective and efficient tuning framework to align large language model with recommendation},
  author={Bao, Keqin and Zhang, Jizhi and Zhang, Yang and Wang, Wenjie and Feng, Fuli and He, Xiangnan},
  booktitle={Proceedings of the 17th ACM Conference on Recommender Systems},
  pages={1007--1014},
  year={2023}
}

@inproceedings{zheng2024adapting,
  title={Adapting large language models by integrating collaborative semantics for recommendation},
  author={Zheng, Bowen and Hou, Yupeng and Lu, Hongyu and Chen, Yu and Zhao, Wayne Xin and Chen, Ming and Wen, Ji-Rong},
  booktitle={2024 IEEE 40th International Conference on Data Engineering (ICDE)},
  pages={1435--1448},
  year={2024},
  organization={IEEE}
}

@article{zhang2023collm,
  title={Collm: Integrating collaborative embeddings into large language models for recommendation},
  author={Zhang, Yang and Feng, Fuli and Zhang, Jizhi and Bao, Keqin and Wang, Qifan and He, Xiangnan},
  journal={arXiv preprint arXiv:2310.19488},
  year={2023}
}

@inproceedings{li2023text,
  title={Text is all you need: Learning language representations for sequential recommendation},
  author={Li, Jiacheng and Wang, Ming and Li, Jin and Fu, Jinmiao and Shen, Xin and Shang, Jingbo and McAuley, Julian},
  booktitle={Proceedings of the 29th ACM SIGKDD Conference on Knowledge Discovery and Data Mining},
  pages={1258--1267},
  year={2023}
}

@inproceedings{xi2024towards,
  title={Towards open-world recommendation with knowledge augmentation from large language models},
  author={Xi, Yunjia and Liu, Weiwen and Lin, Jianghao and Cai, Xiaoling and Zhu, Hong and Zhu, Jieming and Chen, Bo and Tang, Ruiming and Zhang, Weinan and Yu, Yong},
  booktitle={Proceedings of the 18th ACM Conference on Recommender Systems},
  pages={12--22},
  year={2024}
}

@inproceedings{wei2024llmrec,
  title={Llmrec: Large language models with graph augmentation for recommendation},
  author={Wei, Wei and Ren, Xubin and Tang, Jiabin and Wang, Qinyong and Su, Lixin and Cheng, Suqi and Wang, Junfeng and Yin, Dawei and Huang, Chao},
  booktitle={Proceedings of the 17th ACM International Conference on Web Search and Data Mining},
  pages={806--815},
  year={2024}
}

@article{kingma2014adam,
  title={Adam: A method for stochastic optimization},
  author={Kingma, Diederik P},
  journal={arXiv preprint arXiv:1412.6980},
  year={2014}
}

@article{neelakantan2022text,
  title={Text and code embeddings by contrastive pre-training},
  author={Neelakantan, Arvind and Xu, Tao and Puri, Raul and Radford, Alec and Han, Jesse Michael and Tworek, Jerry and Yuan, Qiming and Tezak, Nikolas and Kim, Jong Wook and Hallacy, Chris and others},
  journal={arXiv preprint arXiv:2201.10005},
  year={2022}
}

@inproceedings{yang2023kappahgcn,
  title={$\kappa$hgcn: Tree-likeness modeling via continuous and discrete curvature learning},
  author={Yang, Menglin and Zhou, Min and Pan, Lujia and King, Irwin},
  booktitle={Proceedings of the 29th ACM SIGKDD Conference on Knowledge Discovery and Data Mining},
  pages={2965--2977},
  year={2023}
}

@article{ganea2018hyperbolic,
  title={Hyperbolic neural networks},
  author={Ganea, Octavian and B{\'e}cigneul, Gary and Hofmann, Thomas},
  journal={Advances in neural information processing systems},
  volume={31},
  year={2018}
}

@inproceedings{zheng2017hierarchical,
  title={Hierarchical collaborative embedding for context-aware recommendations},
  author={Zheng, Lei and Cao, Bokai and Noroozi, Vahid and Philip, S Yu and Ma, Nianzu},
  booktitle={2017 IEEE International Conference on Big Data (Big Data)},
  pages={867--876},
  year={2017},
  organization={IEEE}
}

@article{unger2020hierarchical,
  title={Hierarchical latent context representation for context-aware recommendations},
  author={Unger, Moshe and Tuzhilin, Alexander},
  journal={IEEE Transactions on Knowledge and Data Engineering},
  volume={34},
  number={7},
  pages={3322--3334},
  year={2020},
  publisher={IEEE}
}

@article{kunaver2017diversity,
  title={Diversity in recommender systems--A survey},
  author={Kunaver, Matev{\v{z}} and Po{\v{z}}rl, Toma{\v{z}}},
  journal={Knowledge-based systems},
  volume={123},
  pages={154--162},
  year={2017},
  publisher={Elsevier}
}

@incollection{castells2021novelty,
  title={Novelty and diversity in recommender systems},
  author={Castells, Pablo and Hurley, Neil and Vargas, Saul},
  booktitle={Recommender systems handbook},
  pages={603--646},
  year={2021},
  publisher={Springer}
}

@inproceedings{dasgupta2016cost,
  title={A cost function for similarity-based hierarchical clustering},
  author={Dasgupta, Sanjoy},
  booktitle={Proceedings of the forty-eighth annual ACM symposium on Theory of Computing},
  pages={118--127},
  year={2016}
}

@article{yang2024hgformer,
  title={Hgformer: Hyperbolic Graph Transformer for Recommendation},
  author={Yang, Xin and Li, Xingrun and Chang, Heng and Yang, Jinze and Yang, Xihong and Tao, Shengyu and Chang, Ningkang and Shigeno, Maiko and Wang, Junfeng and Yin, Dawei and others},
  journal={arXiv preprint arXiv:2502.15693},
  year={2024}
}

@article{yang2024hyperbolic,
  title={Hyperbolic Fine-tuning for Large Language Models},
  author={Yang, Menglin and Feng, Aosong and Xiong, Bo and Liu, Jihong and King, Irwin and Ying, Rex},
  journal={arXiv preprint arXiv:2410.04010},
  year={2024}
}

@inproceedings{Luo2023cross,
 author = {Luo, Zihan and Huang, Hong and Lian, Jianxun and Song, Xiran and Xie, Xing and Jin, Hai},
 booktitle = {Advances in Neural Information Processing Systems},
 editor = {A. Oh and T. Naumann and A. Globerson and K. Saenko and M. Hardt and S. Levine},
 pages = {79594--79612},
 publisher = {Curran Associates, Inc.},
 title = {Cross-links Matter for Link Prediction: Rethinking the Debiased GNN from a Data Perspective},
 url = {https://proceedings.neurips.cc/paper_files/paper/2023/file/fba4a59c7a569fce120eea9aa9227052-Paper-Conference.pdf},
 volume = {36},
 year = {2023}
}

@article{Masrour_Wilson_Yan_Tan_Esfahanian_2020, title={Bursting the Filter Bubble: Fairness-Aware Network Link Prediction}, volume={34}, url={https://ojs.aaai.org/index.php/AAAI/article/view/5429}, DOI={10.1609/aaai.v34i01.5429}, abstractNote={&lt;p&gt;Link prediction is an important task in online social networking as it can be used to infer new or previously unknown relationships of a network. However, due to the homophily principle, current algorithms are susceptible to promoting links that may lead to increase segregation of the network—an effect known as filter bubble. In this study, we examine the filter bubble problem from the perspective of algorithm fairness and introduce a dyadic-level fairness criterion based on network modularity measure. We show how the criterion can be utilized as a postprocessing step to generate more heterogeneous links in order to overcome the filter bubble problem. In addition, we also present a novel framework that combines adversarial network representation learning with supervised link prediction to alleviate the filter bubble problem. Experimental results conducted on several real-world datasets showed the effectiveness of the proposed methods compared to other baseline approaches, which include conventional link prediction and fairness-aware methods for i.i.d data.&lt;/p&gt;}, number={01}, journal={Proceedings of the AAAI Conference on Artificial Intelligence}, author={Masrour, Farzan and Wilson, Tyler and Yan, Heng and Tan, Pang-Ning and Esfahanian, Abdol}, year={2020}, month={Apr.}, pages={841-848} }

@inproceedings{li2023break,
author = {Li, Zhenyang and Dong, Yancheng and Gao, Chen and Zhao, Yizhou and Li, Dong and Hao, Jianye and Zhang, Kai and Li, Yong and Wang, Zhi},
title = {Breaking Filter Bubble: A Reinforcement Learning Framework of Controllable Recommender System},
year = {2023},
isbn = {9781450394161},
publisher = {Association for Computing Machinery},
address = {New York, NY, USA},
url = {https://doi.org/10.1145/3543507.3583856},
doi = {10.1145/3543507.3583856},
booktitle = {Proceedings of the ACM Web Conference 2023},
pages = {4041–4049},
numpages = {9},
location = {Austin, TX, USA},
series = {WWW '23}
}

@article{wu2022multi,
  title={A multi-objective optimization framework for multi-stakeholder fairness-aware recommendation},
  author={Wu, Haolun and Ma, Chen and Mitra, Bhaskar and Diaz, Fernando and Liu, Xue},
  journal={ACM Transactions on Information Systems},
  volume={41},
  number={2},
  pages={1--29},
  year={2022},
  publisher={ACM New York, NY}
}

@inproceedings{peng2024reconciling,
  title={Reconciling the accuracy-diversity trade-off in recommendations},
  author={Peng, Kenny and Raghavan, Manish and Pierson, Emma and Kleinberg, Jon and Garg, Nikhil},
  booktitle={Proceedings of the ACM Web Conference 2024},
  pages={1318--1329},
  year={2024}
}

@inproceedings{liu2021diversity,
  title={Diversity-promoting deep reinforcement learning for interactive recommendation},
  author={Liu, Yong and Shen, Zhiqi and Zhang, Yinan and Cui, Lizhen},
  booktitle={5th international conference on crowd science and engineering},
  pages={132--139},
  year={2021}
}

@inproceedings{wang2021fully,
  title={Fully hyperbolic graph convolution network for recommendation},
  author={Wang, Liping and Hu, Fenyu and Wu, Shu and Wang, Liang},
  booktitle={Proceedings of the 30th ACM international conference on information \& knowledge management},
  pages={3483--3487},
  year={2021}
}

@inproceedings{chen2022modeling,
  title={Modeling scale-free graphs with hyperbolic geometry for knowledge-aware recommendation},
  author={Chen, Yankai and Yang, Menglin and Zhang, Yingxue and Zhao, Mengchen and Meng, Ziqiao and Hao, Jianye and King, Irwin},
  booktitle={Proceedings of the fifteenth ACM international conference on web search and data mining},
  pages={94--102},
  year={2022}
}

@inproceedings{zhang2022hyperbolic,
  title={A hyperbolic-to-hyperbolic user representation with multi-aspect for social recommendation},
  author={Zhang, Hang and Wang, Hao and Wang, Guifeng and Liu, Jiayu and Liu, Qi},
  booktitle={Proceedings of the 31st ACM International Conference on Information \& Knowledge Management},
  pages={4667--4671},
  year={2022}
}

@inproceedings{ma2021knowledge,
  title={Knowledge-enhanced top-k recommendation in poincar{\'e} ball},
  author={Ma, Chen and Ma, Liheng and Zhang, Yingxue and Wu, Haolun and Liu, Xue and Coates, Mark},
  booktitle={Proceedings of the AAAI conference on artificial intelligence},
  volume={35},
  number={5},
  pages={4285--4293},
  year={2021}
}

\appendix
\section{Proof}
\label{sec:proof}
In this section, we provide a rigorous mathematical proof of Proposition \ref{prop}, demonstrating the unique advantage of the aligning mechanism in the hyperbolic space.

\subsection{Gradient Analysis in Hyperbolic Space.} Suppose $\mathbf{x}$ is a hyperbolic embedding derived from ~\ref{subsec:Hyperbolic Graph Collaborative Filtering}, and $\mathbf{y}$ is the corresponding semantic embedding from ~\ref{subsec: Semantic Representations Generation}, the hyperbolic distance is given by:
\begin{equation}
    d_\mathcal{H}(\mathbf{x}, \mathbf{y}) = \operatorname{arccosh}(z)
\end{equation}
where $z = -\langle \mathbf{x}, \mathbf{y} \rangle_\mathcal{L} = x_0 y_0 - \sum_{i=1}^n x_i y_i$.

For large $\|\mathbf{x}\|$ and $\|\mathbf{y}\|$, given $x_0 = \sqrt{1 + \|\mathbf{x}\|^2}$ and $y_0 = \sqrt{1 + \|\mathbf{y}\|^2}$, we adopt the following approximations for further calculation:
\begin{align}
    x_0 \approx \|\mathbf{x}\|, \quad y_0 \approx \|\mathbf{y}\|, \quad \sqrt{z^2 - 1} \approx z
\end{align}
Thus, the Lorentz product can be approximated as:
\begin{equation}
    z = x_0 y_0 - \sum_{i=1}^n x_i y_i \approx \|\mathbf{x}\|\|\mathbf{y}\|(1 - \cos \theta)
\end{equation}
Let $x_j = \|\mathbf{x}\| \hat{x}_j$ and $y_j = \|\mathbf{y}\| \hat{y}_j$ (where $\hat{x}_j$ represents the direction of $x_j$). The partial derivative of $z$ with respect to $x_j$ is calculated as follows:
\begin{equation}
    \begin{aligned}
        \frac{\partial z}{\partial x_j} &= \frac{x_j}{x_0} y_0 - y_j\\
        &\approx \frac{x_j}{\|\mathbf{x}\|} \|\mathbf{y}\| - y_j\\
        &= \|\mathbf{y}\| (\hat{x}_j - \hat{y}_j)
    \end{aligned}
\end{equation}
Therefore, the gradient norm of $d_\mathcal{H}$ for $\mathbf{x}$ can be approximated by:
\begin{equation}
    \begin{aligned}
        \|\nabla_\mathbf{x} d_\mathcal{H}\| &= \frac{\|\nabla_\mathbf{x} z\|}{\sqrt{z^2 - 1}}\\
        &\approx \frac{\|\nabla_\mathbf{x} z\|}{z}\\
        &= \frac{\|\mathbf{y}\|\|\hat{\mathbf{x}} - \hat{\mathbf{y}}\|}{\|\mathbf{x}\|\|\mathbf{y}\|(1 - \cos \theta)}\\
        &= \frac{\|\hat{\mathbf{x}} - \hat{\mathbf{y}}\|}{\|\mathbf{x}\|(1 - \cos \theta)}
    \end{aligned}
\end{equation}

This result suggests that nodes with large norms experience smaller gradient updates, preserving local structures, while nodes with smaller norms undergo larger updates, refining global hierarchical relationships and effectively capturing hierarchical structures. This aligns with the properties of hyperbolic geometry, where small positional changes near the origin result in significant distance changes, and vice versa. Our dynamic gradient update mechanism is well-suited to this characteristic, enabling precise adjustments and enhancing the accuracy of the updates.

\subsection{Gradient Analysis Euclidean Space.} For comparison, we examine the gradient behavior in Euclidean space. The Euclidean distance is defined as:
\begin{equation}
    f(\mathbf{x}) = d_E(\mathbf{x}, \mathbf{y}) = \|\mathbf{x} - \mathbf{y}\|
\end{equation}
The gradient of $f(\mathbf{x})$ with respect to $\mathbf{x}$ is:
\begin{equation}
    \nabla_\mathbf{x} f(\mathbf{x}) = \frac{\mathbf{x} - \mathbf{y}}{\|\mathbf{x} - \mathbf{y}\|} = \frac{\mathbf{x} - \mathbf{y}}{d_E(\mathbf{x}, \mathbf{y})}
\end{equation}
which has a magnitude of:
\begin{equation}
    \|\nabla_\mathbf{x} f(\mathbf{x})\| = 1
\end{equation}

The constant gradient magnitude demonstrates Euclidean space's uniform update behavior, lacking intrinsic adaptation to hierarchical structures. This fundamental difference in gradient dynamics explains hyperbolic space's superiority for hierarchical representation learning.

\begin{table*}[h]
    \centering
    \caption{Comprehensive model comparison on larger datasets.}
    \begin{tabular}{c|lllll|lllll}
        \toprule
        Datasets & \multicolumn{5}{c|}{Amazon-Books (large)} & \multicolumn{5}{c}{Amazon-CDs (large)}\\ \midrule
        \multicolumn{1}{l|}{Metrics} & Recall@20 & NDCG@20 & Div@20 & H@20 & EPC@20 & Recall@20 & NDCG@20 & Div@20 & H@20 & EPC@20 \\ \midrule
        \multicolumn{1}{l|}{LightGCN} & 0.0242 & 0.0107 & 0.6142 & 11.0605 & 0.8943 & 0.0538 & 0.0248 & 0.0572 & 11.6691 & 0.7869  \\
        \multicolumn{1}{l|}{LightGCL} & 0.0329 & 0.0154 & 0.6547 & 13.4200 & 0.9216 & 0.0677 & 0.0323 & \textbf{0.0641} & 12.9922 & 0.8273 \\
        \multicolumn{1}{l|}{SimGCL} & 0.0318 & 0.0142 & 0.5977 & 9.9114 & 0.8683 & 0.0566 & 0.0262 & 0.0588 & 11.0439 & 0.7649 \\
        \multicolumn{1}{l|}{HGCF} & 0.0545 & 0.0307 & 0.6419 & 13.2311 & 0.9132 & 0.0642 & 0.0364 & 0.0612 & 13.7102 & 0.8072 \\
        \multicolumn{1}{l|}{HICF} & \underline{0.0565} & \underline{0.0319} & \underline{0.6777} & \underline{14.0322} & \underline{0.9347} & \underline{0.0683} & \underline{0.0388} & 0.0629 & \underline{13.7451} & \underline{0.8423} \\
        \multicolumn{1}{l|}{HERec} & \textbf{0.0598} & \textbf{0.0335} & \textbf{0.6832} & \textbf{15.3017} & \textbf{0.9388} & \textbf{0.0712} & \textbf{0.0391} & \underline{0.0632} & \textbf{13.7762} & \textbf{0.8330} \\
        \bottomrule
    \end{tabular}
    \label{tab:large}
\end{table*}

\subsection{Error Bound Analysis}
The approximations used in the proof rely on the assumption that $|x|$ and $|y|$ are large, which is common in high-dimensional hyperbolic embedding spaces. In practice, with embedding dimensions such as 50 (as used throughout our experiments), the norms tend to be sufficiently large for the approximation to hold effectively. To further substantiate this assumption, we calculated the error bound as follows:

The exact Lorentz inner product is given by:
\[
z = \sqrt{1 + |x|^2} \sqrt{1 + |y|^2} - |x||y| \cos \theta,
\]
and is approximated as:
\[
z_\text{approx} = |x||y|(1 - \cos \theta).
\]

Using a Taylor expansion, we have:
\[
\sqrt{1 + |x|^2} = |x| \left( 1 + \frac{1}{2|x|^2} - \frac{1}{8|x|^4} + \cdots \right).
\]

Thus, the absolute error becomes:
\[
|z - z_\text{approx}| = \frac{|y|}{2|x|} + \frac{|x|}{2|y|}.
\]

The relative error in the Lorentz product is bounded by:
\[
\epsilon_z \leq \frac{1}{2} \left( \frac{1}{|x|^2} + \frac{1}{|y|^2} \right) \frac{1}{1 - \cos \theta}.
\]

Additionally, the approximation error in $\sqrt{z^2 - 1} \approx z$ is given by:
\[
\sqrt{z^2 - 1} = z \left( 1 - \frac{1}{2z^2} + \cdots \right),
\]
which implies:
\[
\text{Relative error} \leq \frac{1}{2z^2}.
\]

Combining these results, the overall error bound is:
\[
\text{Relative Error in } |\nabla d_H(x, y)| \leq \frac{1}{2|x|^2(1 - \cos \theta)} + \frac{1}{2z^2}.
\]

For instance, with $|x| = |y| = 5$ and $\theta = \frac{\pi}{2}$, the total error is approximately $2.1\%$, demonstrating high approximation accuracy in our setting.

\begin{table}[t]
    \centering
    \caption{Standard deviation of key metrics across datasets over 10 independent runs.}
    \label{tab:std_metrics}
    \begin{tabular}{lccccc}
    \toprule
    \textbf{Dataset} & \textbf{R@20} & \textbf{N@20} & \textbf{DIV@20} & \textbf{H@20} & \textbf{EPC@20} \\
    \midrule
    Amazon & 0.0010 & 0.0004 & 0.0067 & 0.0035 & 0.0002 \\
    Yelp   & 0.0009 & 0.0004 & 0.0091 & 0.0017 & 0.0014 \\
    Google & 0.0010 & 0.0003 & 0.0077 & 0.0008 & 0.0009 \\
    \bottomrule
    \end{tabular}
\end{table}

\begin{table}[t]
\centering
\caption{Statistics of the Large-scale Datasets}
\label{tab:datasets}
\resizebox{\columnwidth}{!}{
\begin{tabular}{lcccc}
\toprule
Dataset & \# Users & \# Items & \# Interactions & Density \\ \hline
Amazon-Books (Large) & 116,455 & 387,492 & 1,442,011 & 3.20$e^{-5}$ \\
Amazon-CDs (Large) & 61,938 & 88,590 & 779,575 & 1.42$e^{-4}$ \\
\bottomrule
\end{tabular}
}
\label{tab:larger statistics}
\end{table}

\begin{table}[t]
    \centering
    \caption{Direct message passing within hyperbolic space.}
    \label{tab:alternative}
    \begin{tabular}{lccccc}
        \hline
        \textbf{Dataset} & \textbf{R@20} & \textbf{N@20} & \textbf{DIV@20} & \textbf{H@20} & \textbf{EPC@20} \\
        \hline
        Amazon & 0.1721 & 0.1082 & 0.5760 & 12.211 & 0.8773 \\
        Yelp   & 0.1205 & 0.0793 & 0.4005 & 12.321 & 0.8755 \\
        Google & 0.1611 & 0.0988 & 0.3532 & 12.766 & 0.8731 \\
        \hline
    \end{tabular}
\end{table}

\section{More Experimental Results}
\label{sec:more_exp}
In this section, we will show more experimental results to further analyze the flexibility and effectiveness of our \model.

\subsection{Larger-scale Datasets.}
\label{subsec: larger dataset}
To further validate our method within real-world deployment scenarios, we conduct experiments on two large-scale datasets, the statistics of which are summarized in Table \ref{tab:larger statistics}. The Amazon-Books (Large) dataset contains more than ten times the number of users and interactions compared to the standard benchmarks used in previous studies, while the Amazon-CDs (Large) dataset is approximately six times larger.

The results presented in Table \ref{tab:large} align consistently with our earlier findings, demonstrating that HERec achieves superior performance over both hyperbolic and Euclidean baselines. This suggests that the model effectively maintains its predictive accuracy and structural advantages even as the data scale increases significantly.

\subsection{Statistical Significance Testing}
\label{subsec: Statistical Significance Testing}
To ensure the robustness and reliability of our results, we conducted 10 independent runs for each experimental setting and reported the standard deviation for all key metrics. Results are shown in Table \ref{tab:std_metrics}. Across all evaluated datasets and metrics, the proposed model consistently exhibits negligible variance. This high stability demonstrates that our performance gains are resistant to stochastic fluctuations during optimization, confirming the reproducibility and statistical significance of our empirical conclusions.

\subsection{Alternative Hyperbolic Message Passing}
\label{subsec: Alternative hyperbolic message passing}
To bypass the need for tangent space projections, we also explored an alternative message passing strategy by replacing the original hyperbolic graph collaborative filtering mechanism with direct message passing in hyperbolic space. This variant achieves performance comparable to our primary method, results can be found in Table~\ref{tab:alternative}. These findings suggest that direct message passing in hyperbolic space is not only a viable alternative but could also serve as a complementary component within our framework, potentially offering additional flexibility for future extensions. This also underscores the capability of hyperbolic geometries to preserve complex relational hierarchies without relying on Euclidean approximations.

\end{document}